\documentclass{aa}  

\usepackage{graphicx}
\usepackage{txfonts}
\usepackage{lipsum}
\usepackage{subcaption}         % necessary for continued figures, example in section 3
\usepackage{lscape}             % to rotate a single page table, example in appendix.
\usepackage{placeins}           % useful with \FloatBarrier, to keep 
\usepackage[breaklinks=true,colorlinks=true,linkcolor=NavyBlue,citecolor=NavyBlue,filecolor=NavyBlue,urlcolor=NavyBlue]{hyperref}
\usepackage[dvipsnames]{xcolor}

\newcommand{\Msun}{\,\mathrm{M}_\odot}

\newcommand{\km}{\,\mathrm{km}}

\newcommand{\s}{\,\mathrm{s}}

\newcommand{\dy}{\,\mathrm{d}}
\newcommand{\gcminvcb}{\,\mathrm{g}\,\mathrm{cm}^{-3}}
\newcommand{\kmsinv}{\,\mathrm{km}\,\mathrm{s}^{-1}}
\newcommand{\K}{\,\mathrm{K}}
\newcommand{\erg}{\,\mathrm{erg}}
\newcommand{\ergsinv}{\,\mathrm{erg}\,\mathrm{s}^{-1}}
\newcommand{\ang}{\,\mathrm{\AA}}

\newcommand{\mbh}{m_{\mathrm{BH}}}
\newcommand{\mwd}{m_{\mathrm{WD}}}
\newcommand{\rwd}{r_{\mathrm{WD}}}
\newcommand{\tdyn}{t_{\mathrm{dyn}}}
\newcommand{\tcross}{t_{\mathrm{cross,WD}}}
\newcommand{\rt}{r_{\mathrm{t}}}
\newcommand{\rp}{r_{\mathrm{p}}}
\newcommand{\rs}{r_{\mathrm{s}}}
\newcommand{\munb}{m_{\mathrm{unbound}}}
\newcommand{\mproj}{m_{\mathrm{proj,10^{\circ}}}}
\newcommand{\funb}{f_{\mathrm{unbound}}}
\newcommand{\EKunb}{E_{\mathrm{K,unbound}}}

\newcommand\Tstrut{\rule{0pt}{2.9ex}}         % "top" strut
\newcommand\Bstrut{\rule[-1.2ex]{0pt}{0pt}}   % "bottom" strut
\newcommand\TBstrut{\Tstrut\Bstrut}           % "top and bottom" strut

\newcommand{\code}[1]{\texttt{#1}}
\newcommand{\voned}{\code{V1D}}
\newcommand{\mesa}{\code{MESA}}
\newcommand{\cmfgen}{\code{CMFGEN}}
\newcommand{\arepo}{\code{AREPO}}
\newcommand{\longpol}{\code{LONG\_POL}}
\newcommand{\sedona}{\code{SEDONA}}

\newcommand{\iso}[2]{\ensuremath{^{#1}\rm{#2}}}
\def\ctwelve{\iso{12}C}
\def\osixteen{\iso{16}O}
\def\nifs{\iso{56}Ni}
\def\cofs{\iso{56}Co}

\def\caiidoub{[Ca\two]\,$\lambda\lambda$\,$7291,\,7323$}
\def\oidoub{[O\one]\,$\lambda\lambda$\,$6300,\,6364$}

\def\one{{\,\sc i}}
\def\two{{\,\sc ii}}
\def\three{{\,\sc iii}}

\begin{document}

   \title{Asymmetrical thermonuclear supernovae triggered by the tidal disruption of white dwarfs}
   \titlerunning{Supernovae from WD TDEs}
   
   \subtitle{}

   \author{Pavan Vynatheya\inst{1,5}\fnmsep\thanks{corresponding authors},
        Luc Dessart\inst{2}\fnmsep$^{\star}$,
        Taeho Ryu\inst{3,4,5}
        \and Rüdiger Pakmor\inst{5}
        }
    \authorrunning{Vynatheya et al.}

   \institute{Canadian Institute for Theoretical Astrophysics, University of Toronto, 60 St George St, Toronto, ON M5S 3H8, Canada
            \and Institut d’Astrophysique de Paris, CNRS-Sorbonne Université, 98 bis boulevard Arago, F-75014 Paris, France            
            \and JILA, University of Colorado and National Institute of Standards and Technology, 440 UCB, Boulder, 80308 CO, USA
            \and Department of Astrophysical and Planetary Sciences, 391 UCB, Boulder, 80309 CO, USA
            \and Max-Planck-Institut für Astrophysik, Karl-Schwarzschild-Straße 1, 85748 Garching bei München, Germany}

   \date{Received XX, YY}

% \abstract{}{}{}{}{}
% 5 {} token are mandatory
 
  \abstract{In a dense star cluster core, a tidal disruption event (TDE) of a white dwarf (WD) can occur if the WD passes within the tidal radius of an intermediate-mass black hole (IMBH). Very close encounters cause extreme tidal compression in the WD, raising temperatures enough to induce runaway fusion and produce a thermonuclear supernova (SN). Using the hydrodynamics code \arepo\ augmented with a 55-isotope nuclear reaction network, we performed high-resolution simulations of the TDE of a $0.6 \Msun$ C/O WD by a $500 \Msun$ IMBH for different values of the scaled impact parameter $b$ (i.e., the ratio of periapsis distance to tidal radius). Closer encounters produce combined TDE+SN events, with a partial burning of \ctwelve\ and \osixteen\ into heavier isotopes -- the \nifs\ fractions of the disrupted WD material vary from 1\% at $b = 0.19$ to 82\% at $b = 0.10$, while wider ones ($b \gtrsim 0.20$) lead to standard TDEs. In all cases, the material away from the denser regions remains unburnt, spanning a wide range of radial velocities. Such WD TDEs also exhibit a central cavity, wherein little material is found below a radial velocity of several $1000 \kmsinv$. We also performed 1D and 2D radiative-transfer calculations for these WD-TDEs using the codes \cmfgen\ and \longpol, respectively, covering epochs from a few days to one hundred days. We recover the typical rise times and peak luminosities of SNe Ia, but with an extremely strong viewing-angle dependence of both light curves and spectra. At nebular times, isolated strong emission lines such as \caiidoub\ may appear both displaced and skewed by many $1000 \kmsinv$ -- such extreme offsets are harder to identify at earlier times due to optical depth effects and line overlap. WD TDEs may produce a diverse set of transients with extreme asymmetry and peculiar composition.}
  

   \maketitle
   \nolinenumbers

%%%%%%%%%%%%%%%%%%%%%%%%%%%%%%%%%%%%%%%%%%%%%%%%%%%%%%%%%%%%%%
\section{Introduction} \label{sec:intro}

When a stellar object approaches a supermassive black hole (SMBH) in a galactic center, it can be either partially or fully disrupted due to extreme tidal forces. Such an event is called a tidal disruption event \citep[TDE;][]{1988Natur.331..687H,1988Natur.333..523R}. So far, a few hundred TDE candidates have been observed \citep[e.g.,][]{2023arXiv231016879W}. The new Vera C. Rubin Observatory (formerly Large Synoptic Survey Telescope; \citealt{2019ApJ...873..111I}) is especially promising for transient astronomy and is predicted to detect thousands of TDEs \citep{2020ApJ...890...73B,2024A&A...690A.384S,2025A&A...701A.142M}.

Although their event rates are predicted to be lower \citep[e.g.,][]{2016ApJ...823..113P,2023ApJ...953..141Y}, TDEs can also occur in dense star clusters around stellar-mass BHs (SBHs; masses $\lesssim 10^2 \Msun$; e.g., \citealt{2016ApJ...823..113P,2024A&A...685A..45V}) which are the end products of massive stellar evolution, or elusive intermediate-mass BHs (IMBHs; masses $\sim 10^2$ -- $10^5 \Msun$; e.g., \citealt{2023ApJ...953..141Y,2025ApJ...980L..22C}), the focus of this work, which are speculated to reside in the centers of massive star clusters \citep[e.g.,][]{2020ARA&A..58..257G}. Recently, \citet{2024Natur.631..285H} made the first conclusive discovery of an IMBH of mass $\sim 8000 \Msun$ in the center of $\omega$ Centauri by observing the velocities of selected stars near the center. Observations of TDEs of white dwarfs (WDs) would provide conclusive evidence of the presence of IMBHs \citep[e.g.,][]{2011ApJ...726...34C,2020SSRv..216...39M,2023ApJ...955...46G} since WDs cannot be tidally disrupted by SMBHs and instead plunge directly into them without observational signatures. This is because an SMBHS's Schwarzchild radius, $\rs = 2 G \mbh / c^2$, exceeds a WD's tidal radius, $\rt = (\mbh/\mwd)^{1/3} \rwd$, where $\mbh$, $\mwd$, and $\rwd$ are the mass of the BH, mass of the WD, and the radius of the WD, respectively. 

In this study, we focus on tidal encounters between WDs and IMBHs (henceforth referred to as WD TDEs). Observations of these transients have been lacking due to their expected rarity \citep[e.g.,][]{2020SSRv..216...39M,2023ApJ...955...46G}, although some studies \citep[e.g.,][]{2013ApJ...779...14J,2019MNRAS.487.2505K,2025arXiv250925877L,2026MNRAS.546ag005E} attribute unusual TDEs with strong flares over short timescales to WD TDEs, among others. WD TDEs have also been of interest theoretically, with hydrodynamics studies of WD-IMBH encounters \citep[e.g.,][]{2009ApJ...695..404R,2017ApJ...841..132L,2025arXiv250803463M} being conducted. In particular, \citet{2009ApJ...695..404R} performed a suite of smoothed-particle hydrodynamics (SPH) simulations of WD TDEs with a nuclear reaction network and showed that they can result in thermonuclear supernovae (SNe). This occur only if the tidal compression is sufficiently large for temperatures to be optimal for runaway nuclear reactions. \citet{2016ApJ...819....3M} followed this up by modeling the line-of-sight-dependent light curves and spectra, during the photospheric phase ($\lesssim 40 \dy$), of one of these encounters (that of a $0.6 \Msun$ C/O WD) using 3D Monte Carlo radiative transfer. They found that it observationally resembles type Ia SNe, and shows large line-profile wavelength shifts depending on viewing angle.

In our work, we perform high-resolution simulations of WD TDEs for different encounter impact parameters using moving-mesh hydrodynamics, and model their light curves and spectra using 1D and 2D radiative transfer codes. Our main motivation is to understand how the ejecta and their observational signatures transition from a standard TDE to a thermonuclear SN depending on how close the WD approaches the IMBH. Consequently, we show that WD TDEs can describe a wide range of observable transients that are highly dependent not only on the viewing angle but also on the encounter impact parameter. 

The paper is structured as follows. Sections \ref{sec:hydro} and \ref{sec:result} present the methods and results, respectively, for the hydrodynamics simulations of close WD-IMBH encounters. Sections \ref{sect_rt}
 and \ref{sect_longpol} outline and detail the 1D and 2D radiative transfer modeling, respectively, of the unbound ejecta. Finally, Section \ref{sec:conclude} provides a conclusion.
 
%%%%%%%%%%%%%%%%%%%%%%%%%%%%%%%%%%%%%%%%%%%%%%%%%%%%%%%%%%%%%%
\section{Hydrodynamics and stellar physics} \label{sec:hydro}

We employed the 3D hydrodynamics code \arepo\ \citep{2010MNRAS.401..791S,2016MNRAS.455.1134P,2020ApJS..248...32W} to simulate the tidal disruption of a WD by an IMBH. In contrast to the two popular schemes -- the Eulerian finite-volume static grid and the Lagrangian smoothed particle hydrodynamics (SPH) -- \arepo\ is an arbitrary Lagrangian-Eulerian moving-mesh code that solves the fluid equations on an unstructured Voronoi grid moving (approximately) with the fluid bulk velocity. It has the advantages of both methods; for example, the lack of need for artificial viscosity and superior shock treatment of static grid codes, and the inherent adaptive resolution and Lagrangian invariance of SPH codes. 

Although originally written for cosmological simulations, \arepo\ can be repurposed for stellar astrophysics by implementing a stellar equation of state (EoS). We used the \texttt{Helmholtz} EoS \citep{2000ApJS..126..501T} that describes, along with an ionized ideal gas, an arbitrarily degenerate, arbitrarily relativistic electron-positron gas, and includes radiation from a black-body with the local gas temperature. Furthermore, we employed a nuclear reaction network (NRN) using reaction rates from the \texttt{JINA REACLIB} database \citep{2010ApJS..189..240C} and weak interaction rates \citep{2001ADNDT..79....1L} that includes 55 isotope species: $^{1}_{0}$n, $^{1}_{1}$p (or $^{1}$H), $^{4}$He, $^{11}$B, $^{12,13}$C, $^{13-15}$N, $^{15-17}$O, $^{18}$F, $^{19-22}$Ne, $^{22,23}$Na, $^{23-26}$Mg, $^{25-27}$Al, $^{28-30}$Si, $^{29-31}$P, $^{31-33}$S, $^{33-35}$Cl, $^{36-39}$Ar, $^{39}$K, $^{40}$Ca, $^{43}$Sc, $^{44}$Ti, $^{47}$V, $^{48}$Cr, $^{51}$Mn, $^{52,56}$Fe, $^{55}$Co, $^{56,58,59}$Ni. \citet{2012MNRAS.424.2222P} describes this implementation in the SPH code \texttt{GADGET} \citep{2005MNRAS.364.1105S}, which has since been extended to \arepo\ \citep[e.g.,][]{2013ApJ...770L...8P}. After each hydrodynamic timestep, the NRN is integrated, and the subsequent compositional and (internal) energy changes for each cell are updated. Nuclear reactions are computed only for cells with temperatures $T > 2\times10^7 \K$. This is reasonable since carbon burning requires temperatures $T \gtrsim 5\times10^8 \K$. This is a major improvement over \citet{2016ApJ...819....3M}, whose 7-species NRN provides a much less accurate description of the energy release and the composition of the ejecta, both of which are important for radiative transfer calculations and the resulting observables.

In our simulations, cell refinement (or de-refinement) was triggered when the mass within a cell deviated by more than a factor of two from the mean cell mass, with additional refinement occurring when adjacent cells had a volume difference of a factor of ten or more. The minimum gravitational softening length of the fluid cells was set to one-tenth of the smallest fluid cell in the simulation. We did not include magnetic fields in our simulations.

\subsection{White dwarf}
Similarly to \citet{2012MNRAS.424.2222P}, we constructed the density and pressure profiles of a $0.6 \Msun$, nonrotating C/O WD (50\% \ctwelve\ and 50\% \osixteen)\footnote{C/O ratios depend on masses and metallicities of initial stars, and impact future nuclear burning.}, with central density $\rho_{\mathrm{c, WD}} = 3.7 \times 10^6 \gcminvcb$ and radius $\rwd = 8400 \km$. This was achieved by assuming an initial temperature of $T = 5\times10^5\,\mathrm{K}$ and solving for hydrostatic equilibrium.  We then mapped these 1D profiles to a 3D \arepo\ mesh comprised of almost cubical cells of nearly equal masses arranged in HEALPix shells \citep{2005ApJ...622..759G,2017A&A...599A...5O}. The choice of WD mass was motivated by the fact that the WD mass distribution peaks at $\sim 0.6 \Msun$ \citep[e.g.,][]{2007MNRAS.375.1315K,2011ApJ...730..128T,2013ApJS..204....5K}.

We set an ambient density of $10^{-10} \gcminvcb$ for the ``vacuum'' cells surrounding the WD. These cells are much larger than the WD cells and span a box ten times larger than the farthest distance between the WD and the IMBH. The volume refinement criterion affects the vacuum cells in the immediate vicinity of the WD resulting in a smoothing of its surface, but this effect is relatively negligible. The mass refinement criterion does not apply to these cells.

Introducing artificial damping to remove spurious velocities, we then relaxed the WD for ten sound-crossing timescales $\displaystyle \tcross = 2 \sum_{i=1}^{n} c_{\mathrm{s}}(r_i)\ (r_i - r_{i-1}) = 12.7 \s$, where $r_0 = 0$, $r_n = \rwd$, and $c_{\mathrm{s}}(r_i)$ is the speed of sound at radial distance $r_i$. Figure \ref{fig:density_wd} shows the density profiles of the initial and relaxed WDs. Even after $10\,\tcross$, the density profiles remain roughly the same. We chose a WD cell resolution of $10^{-7}~\Msun/\mathrm{cell}$ by performing TDE test runs with different resolution values, namely $10^{-6}~\Msun/\mathrm{cell}$, $5\times10^{-6}~\Msun/\mathrm{cell}$ and $10^{-7}~\Msun/\mathrm{cell}$. We found that the saturation levels of massive isotope production after periapsis passage, among other quantities, converge for cell resolutions of $5\times10^{-6}~\Msun/\mathrm{cell}$ and $10^{-7}~\Msun/\mathrm{cell}$.

\subsection{Black hole}
In our simulations, a $500 \Msun$ IMBH was included as a non-rotating object that interacts only gravitationally with fluid cells. For our chosen WD and BH masses, the relevant length scales are as follows:
\begin{itemize}
    \item Schwarzschild radius $\rs = 2 G \mbh / c^2 \approx 1500 \km$.
    \item Tidal radius $\rt = (\mbh/\mwd)^{1/3} \rwd \approx 8 \times 10^4 \km \approx 55 \rs$.
\end{itemize}
In this work, we explored full TDEs (FTDEs) with periapsis distances of $\rp \approx 5.5$ -- $ 11 \rs$, where relativistic effects are expected to be only mildly important for most of the encounters. Consequently, we adopted Newtonian gravity in our simulations and set the gravitational softening length of the BH to the size of the smallest fluid cell.

\subsection{Initial conditions}
We placed the relaxed WD and the IMBH on parabolic orbits with varying impact parameters $b = \rp/\rt$ between 0.10 and 0.20, where $\rp$ is the periapsis (closest approach) distance. We considered small increments in $b$ from 0.15 to 0.20 as this range has a steep transition in massive isotope (specifically $^{56}$Ni) production (see Section \ref{sec:result}). The setup is identical to the one described in \citet{2024A&A...685A..45V}, except that the star is replaced by a WD.

The choice of IMBH mass was partly motivated by \citet{2009ApJ...695..404R} (runs 8 and 9 of their Table 1), who calculated that a TDE at $b = 0.20$ with the same WD and IMBH masses would lead to a SN explosion.
The simulation parameters of our run are as follows:
\begin{itemize}
    \item C/O WD mass $\mwd = 0.6 \Msun$, resolved with $\sim 6\times10^6$ cells.
    \item IMBH mass $\mbh = 500 \Msun$, a point mass interacting only gravitationally (Newtonian).
    \item Eight values of the impact parameter $b = \rp/\rt$: 0.20, 0.19, 0.18, 0.17, 0.16, 0.15, 0.12, 0.10.
\end{itemize}

We ran each simulation for 200 dynamical timescales at $\rt$: $\tdyn = [\rt^3 / G (\mbh + \mwd)]^{1/2} = 2.7 \s$\footnote{Since $\mbh \gg \mwd$, this is also the same as the WD dynamical timescale $r_{\mathrm{dyn,WD}} = (\rwd^3 / G \mwd)^{1/2}$.} (periapsis passage occurs at $\sim 15 \tdyn$). Thus, the simulations were terminated $\sim 500 \s$ after periapsis passage. In addition to these eight simulations, we simulated one other TDE with $b = 0.15$ without the NRN enabled to evaluate the impact of nuclear burning on the dynamical evolution of the disruption.

%%%%%%%%%%%%%%%%%%%%%%%%%%%%%%%%%%%%%%%%%%%%%%%%%%%%%%%%%%%%%%
\section{Results from hydrodynamics} \label{sec:result}

\begin{figure*}
    \centering
    \includegraphics[width=\textwidth]{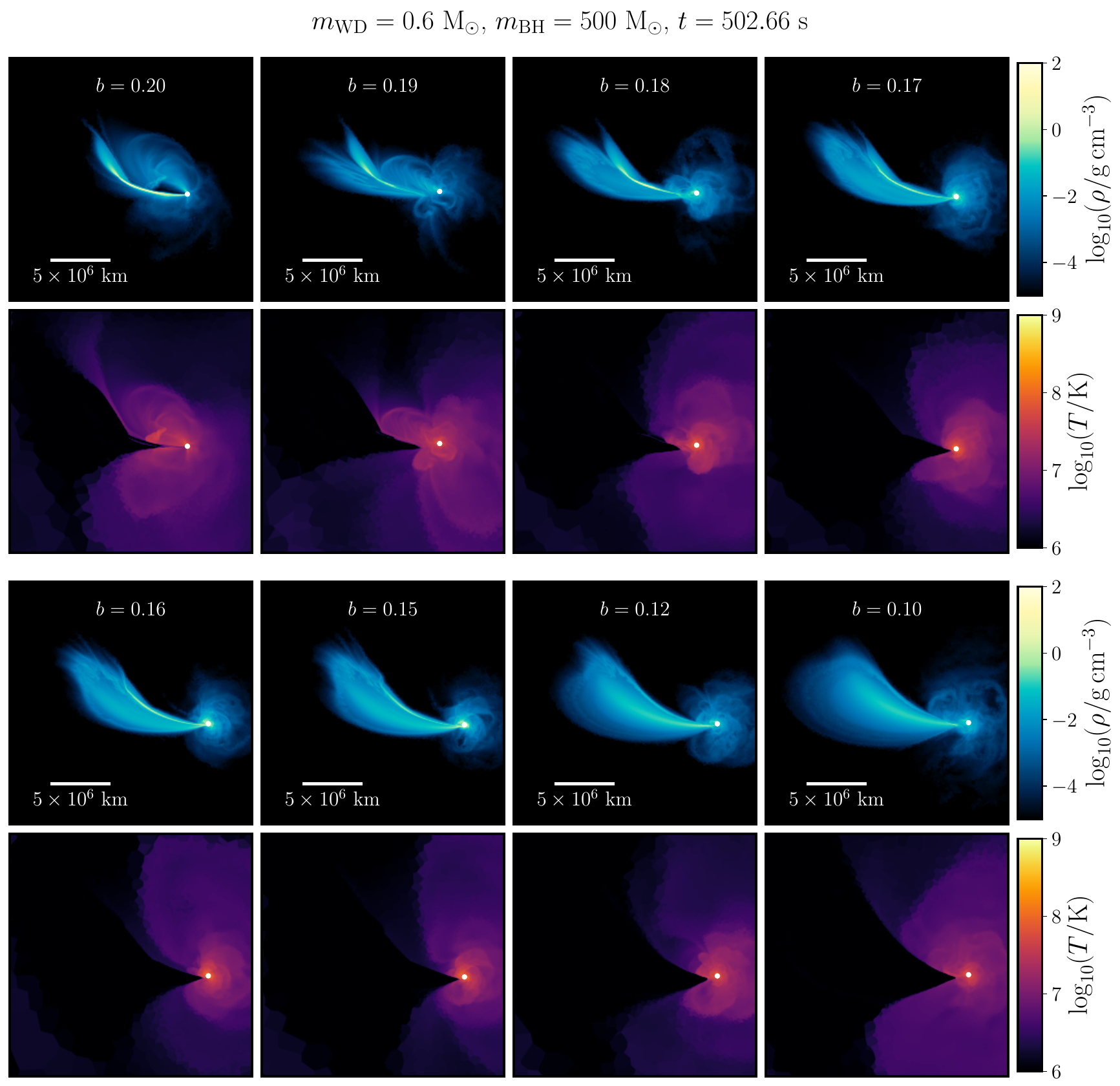}
    \caption{Snapshots of TDEs of $0.6 \Msun$ WDs, due to $500 \Msun$ IMBHs (white points) for eight different values of impact parameter $b$ at $\sim 500 \s$. The top two (bottom two) panels show slices of densities and temperatures for $b = 0.20, 0.19, 0.18, 0.17$ ($b = 0.16, 0.15, 0.12, 0.10$), respectively. As $b$ decreases, the ejecta are more spread out and increasingly unlike a standard TDE, indicating runaway nuclear burning (e.g., the case of $b = 0.10$). The masses of the unbound ejecta are shown in Table \ref{tab:mass_energy}.}
    \label{fig:collage_quants}
\end{figure*}

\begin{figure*}
    \centering
    \includegraphics[width=\textwidth]{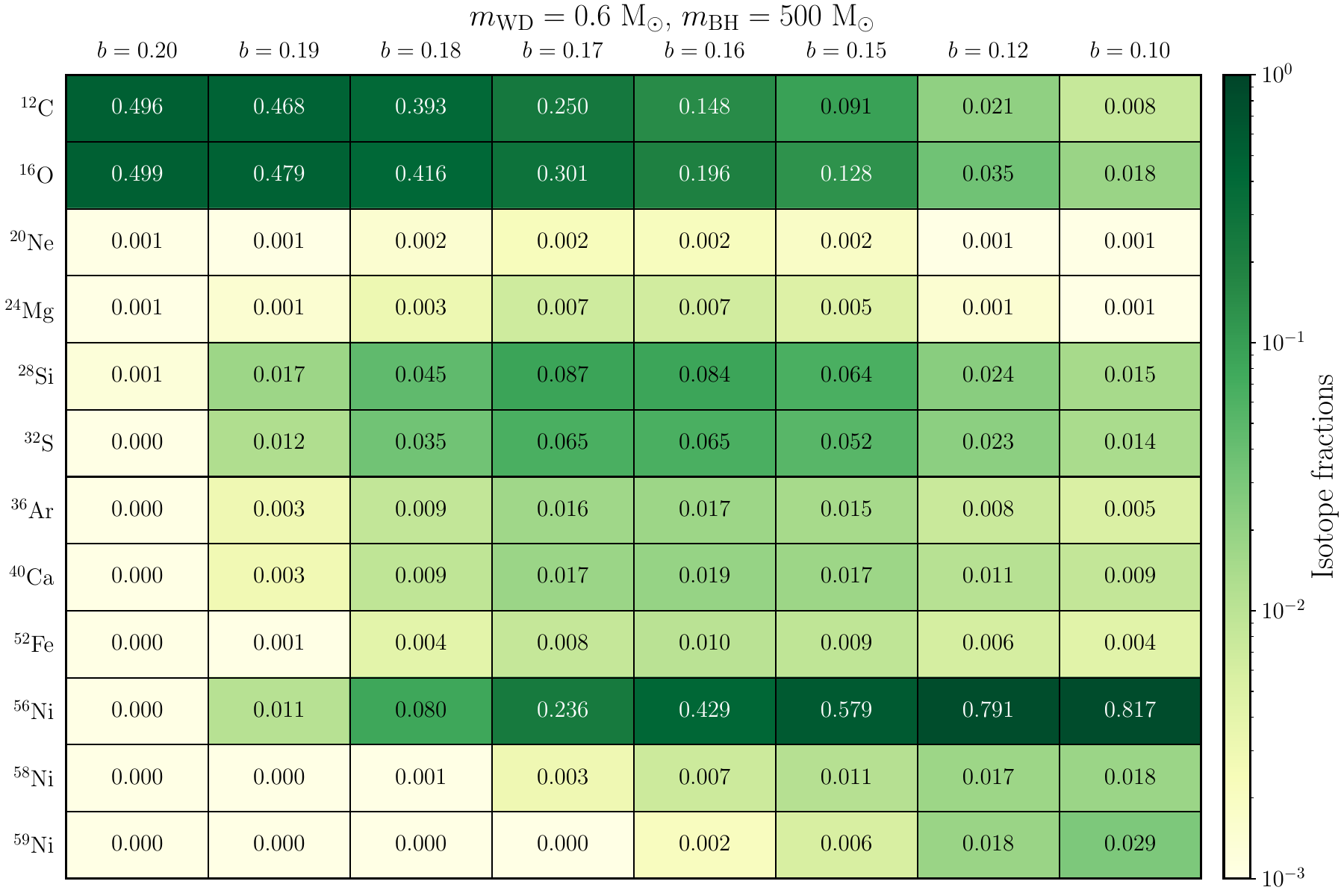}
    \caption{Isotope mass fractions (those with abundances $> 10^{-3}$) of the WD ejecta after periapsis passage for different values of $b$. As $b$ decreases, the fractions of \ctwelve\ and \osixteen\ after TDE decrease due to more nuclear burning, and the fraction of \nifs\ produced increases quickly. Intermediate-mass isotopes (Ne to Ca) first increase in fraction with decreasing $b$, and then decrease due to more favorable conditions for massive isotope (e.g., \nifs) production.}
    \label{fig:nuc_tab}
\end{figure*}

\subsection{Overview}
Since the masses of the WD and the IMBH were fixed in all eight simulations, our results highlight the effect of the impact parameter $b$. As a general overview, smaller impact parameters result in more significant nuclear burning, and at very small $b$, the nuclear burning energy completely dominates the entire energy budget. The essence of this statement is captured in Figure \ref{fig:collage_quants}, illustrating in-plane slices of densities and temperatures of WD-IMBH encounters for different values of $b$ at $\sim 500 \s$ after periapsis passage. The density snapshots clearly show that smaller values of $b$ result in the ejecta being more spread out. The $b = 0.20$ encounter results in a thin stream of tidally disrupted WD material, roughly half of it falling toward the IMBH and the other half being ejected. On the other extreme, the $b = 0.10$ encounter leads to the ejected WD material spreading out, not exhibiting a thin tidal stream (see Table \ref{tab:mass_energy} for unbound ejecta masses). The fanning out of ejecta is attributed to the energy injected into them due to exothermic nuclear fusion reactions. This is evidenced by the two $b = 0.15$ encounter simulations, one with the NRN enabled and the other without: while the latter displays spread-out ejecta (Figure \ref{fig:collage_quants}), the former shows a tidal stream (not shown, but morphologically similar to the $b = 0.20$ case).

\begin{figure*}
    \centering
    \includegraphics[width=\textwidth]{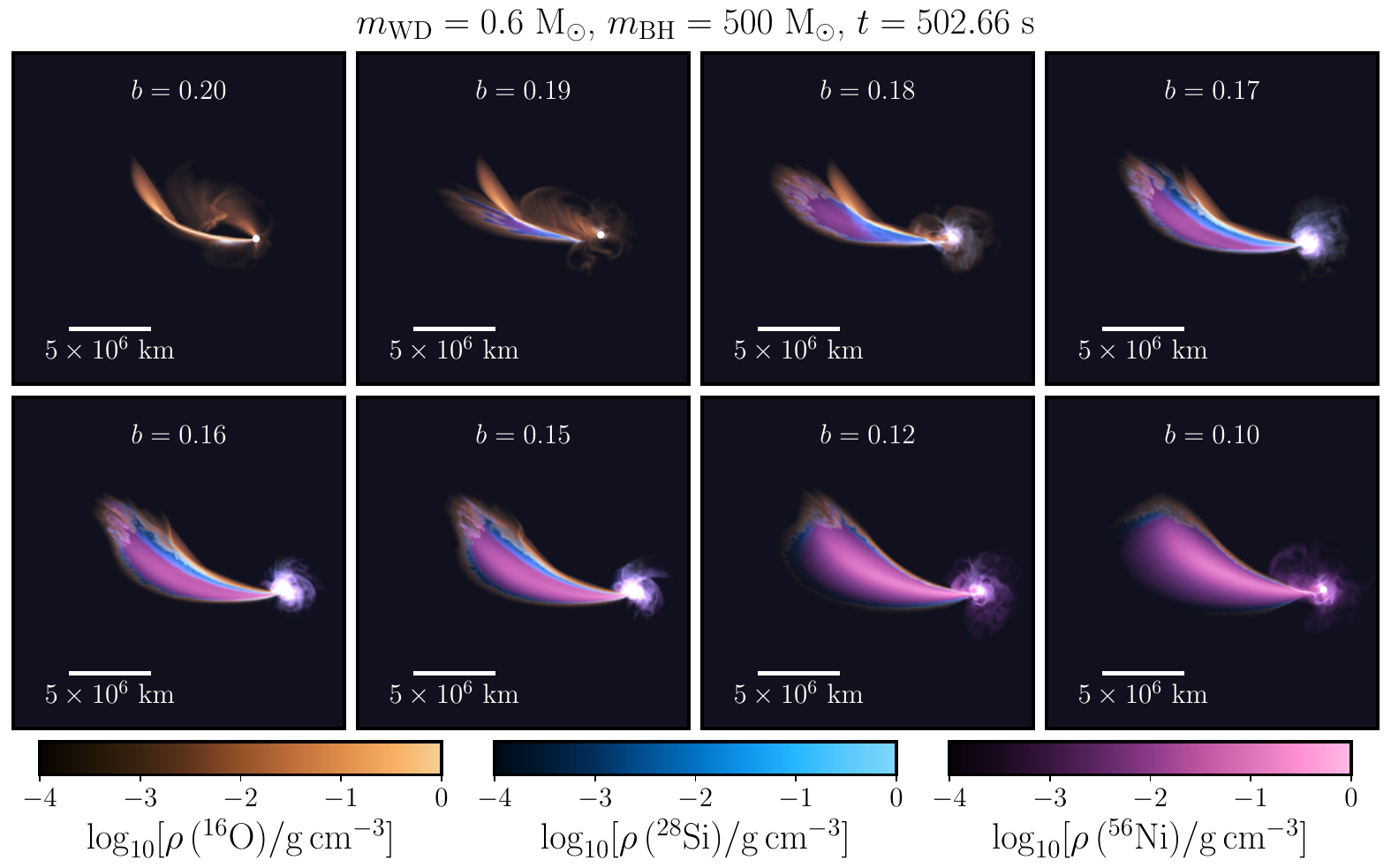}    \caption{Snapshots of slices of \osixteen, $^{28}$Si and \nifs\ densities, respectively, similar to Figure \ref{fig:collage_quants}. As $b$ decreases, a higher abundance of \nifs\ is produced, at the expense of \osixteen\ (and \ctwelve), due to runaway nuclear burning. Abundances of $^{28}$Si (and other isotopes from Ne to Ca) initially increase with decreasing $b$, peaking around $b \sim 0.16$ -- $0.17$, and decrease after (see also Figure \ref{fig:nuc_tab}). When nuclear burning is significant, the inner, intermediate, and outer regions of the ejecta (relative to the centers of the plumes of debris) are dominated by \nifs, $^{28}$Si (also other intermediate isotopes), and \osixteen\ (also \ctwelve), respectively. The transition from a standard TDE ($b = 0.20$) to TDE+SN is clear.}
    \label{fig:collage_nuc}
\end{figure*}

\subsection{Nuclear abundances}
As encounters become stronger, i.e., as $b$ decreases, maximum temperatures of the tidally-compressed degenerate material increase, resulting in greater amounts of \ctwelve\ and \osixteen\ being converted to heavier isotopes. Runaway fusion occurs at periapsis passage, when the tidal compression is strongest. Thus, the timescale of nuclear fusion is comparable to $\tdyn = 2.7 \s$ (see also Section \ref{sec:hydro}). After this brief period, no fusion occurs, and nuclear fractions stabilize.

Figure \ref{fig:nuc_tab} provides a comprehensive table of nuclear mass fractions of the WD ejecta after periapsis passage. Only the isotopes with abundances $> 10^{-4}$ are listed here. The trends of some important isotopes are as follows:
\begin{itemize}
    \item \ctwelve\ and \osixteen\ fractions range from $\sim 50\%$ when $b = 0.20$ (virtually no fusion) to $\sim 1$ -- $2\%$ when $b = 0.10$. More than half of their initial fractions are depleted when $b = 0.16$ -- $0.17$.
    \item $^{28}$Si and $^{32}$S fractions peak at $\sim 6$ -- $9\%$ when $b = 0.16$ -- $0.17$ and then decrease to $\sim 1.5\%$ when $b = 0.10$.
    \item $^{36}$Ar and $^{40}$Ca fractions peak at $\sim 2\%$ when $b = 0.16$ and then decrease to $< 1\%$ when $b = 0.10$.
    \item \nifs\ fractions increase from $\sim 1\%$ when $b = 0.19$ (basically 0 when $b = 0.20$) to $\sim 82\%$ when $b = 0.10$. The largest relative increase occurs in the range $b = 0.16$ -- $0.19$ when the \nifs\ mass fraction increases by a factor of $40$.
    \item $^{58}$Ni and $^{59}$Ni fractions also increase up to $\sim 2$ -- $3\%$ when $b = 0.10$.
\end{itemize}

As expected, deeper encounters lead to further nuclear burning due to increasingly favorable conditions. Consequently, with decreasing b, \ctwelve\ and \osixteen\ fractions decrease, intermediate isotope ($^{28}$Si, $^{32}$S, $^{36}$Ar, $^{40}$Ca, etc.) fractions first increase and then decrease, and \nifs\ (and other massive isotopes) fraction increase. Such properties are standard of thermonuclear fusion and lead to ejecta with relative abundances between iron-group elements (IGEs), intermediate-mass elements (IMEs), and unburnt C/O that are roughly in line with SN Ia models, whether one considers Chandrasekhar mass or sub-Chandrasekhar mass WDs (see, for example, \citealt{2013MNRAS.429.1156S}). The main differences are the relatively lower abundances of IMEs (either material remains unburnt or tends to burn all the way to IGEs), the wider range of values for the \nifs\ fraction (from as low as zero at $b = 0.20$ up to $\sim 82\%$ at $b = 0.10$, which extends beyond the \nifs\ fraction of 61\% in Chandrasekhar mass model DDC0 or 73\% of sub-Chandrasekhar mass model SCH7p0 of \citealt{2017MNRAS.470..157B}). This extreme range thus covers from explosions more typical of stripped-envelope (or perhaps ultra-stripped) SNe of Type Ic (C/O-rich debris for $b = 0.19$), to faint SNe Ia with a low \nifs\ ejected mass (e.g., 91bg-like events; \citealt{1992AJ....104.1543F}) up to the most IGE-dominated SNe Ia as inferred for 91T-like events (e.g., \citealt{1992ApJ...384L..15F};  \citealt{1994ApJ...434L..19S}; \citealt{2024ApJS..273...16P}).

Figure \ref{fig:collage_nuc} is a collage of in-plane slices of isotope densities of \osixteen, $^{28}$Si, and \nifs\ at $\sim 500 \s$. When $b = 0.20$, most of the ejecta are composed of \ctwelve\ (not shown) and \osixteen, and the fractions of \nifs\ increase with decreasing $b$. The onion-like structure of the ejecta is quite evident, with \nifs, $^{28}$Si, and \osixteen\ dominating the inner, intermediate, and outer regions relative to the centers of the debris plumes. This onion-like structure also holds for other isotopes, with more massive isotopes residing in inner shells. For example, \ctwelve, similar to \osixteen, is found in the outer regions; $^{32}$S, $^{36}$Ar, $^{40}$Ca, etc., as $^{28}$Si, are present in the intermediate regions; and so on. It is important to note that, since the centers of the debris plumes are not at rest but move away from the black hole (unlike standard SNe Ia that are point explosions), this stratification is not present in the radial velocity space; i.e., ejecta at all velocities are composed of material of different compositions, contrary to standard SNe Ia.

\begin{figure*}
    \centering
    \includegraphics[width=\textwidth]{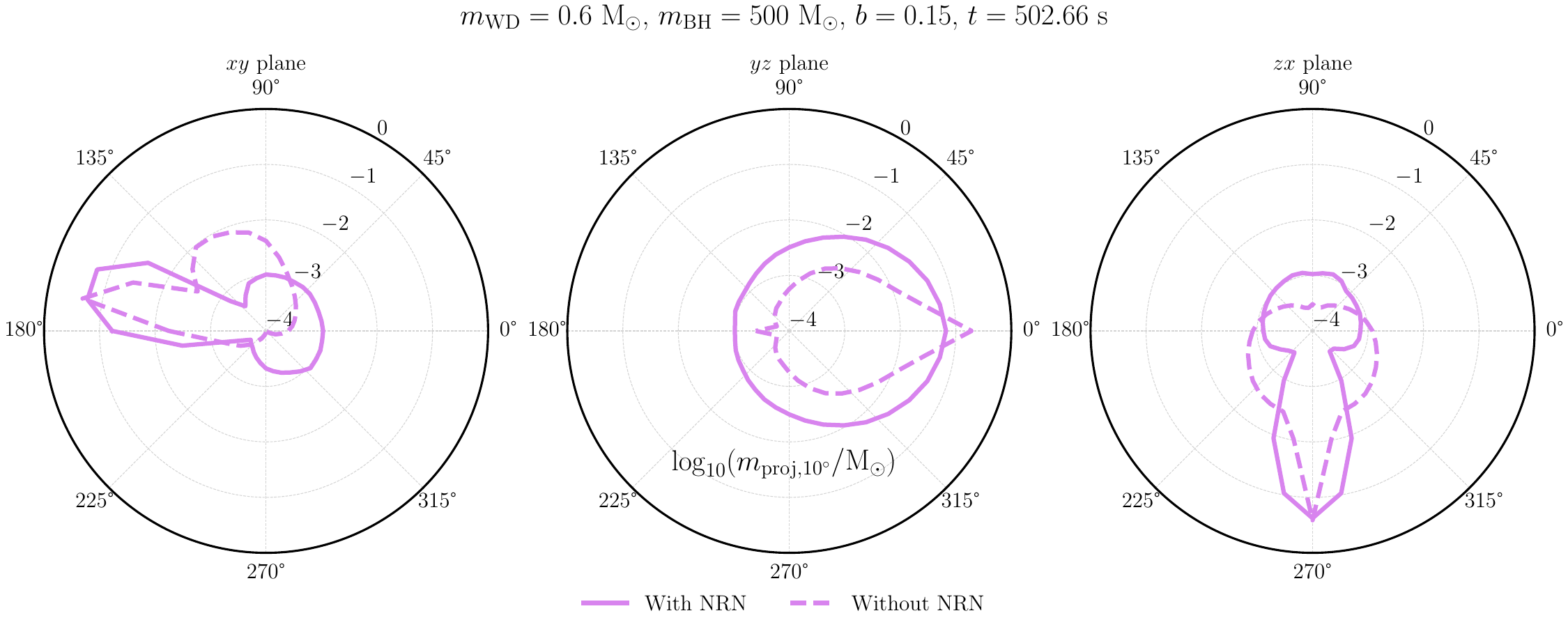}
    \caption{Polar plots of plane-projected masses within wedges of opening angle $10^{\circ}$, $\mproj$, unbound from the IMBHs at $\sim 500 \s$. The solid and dashed lines represent the $b = 0.15$ TDE when the NRN is enabled and disabled, respectively. The three subplots represent three perpendicular planes, $xy$, $yz$, and $zx$, passing through the IMBHs (plot centers). The cylindrical angles are defined relative to the $x$, $y$, and $z$ axes, respectively. Note that the values denote the sum of the projected column masses for each angle. A standard TDE ensues when the NRN is disabled, while nuclear fusion occurs when the NRN is enabled. Due to energy injected from nuclear reactions, the ejected material, in this case, spreads out more.}
    \label{fig:mass_asym_0.15}
\end{figure*}

\begin{figure*}
    \centering
    \includegraphics[width=\textwidth]{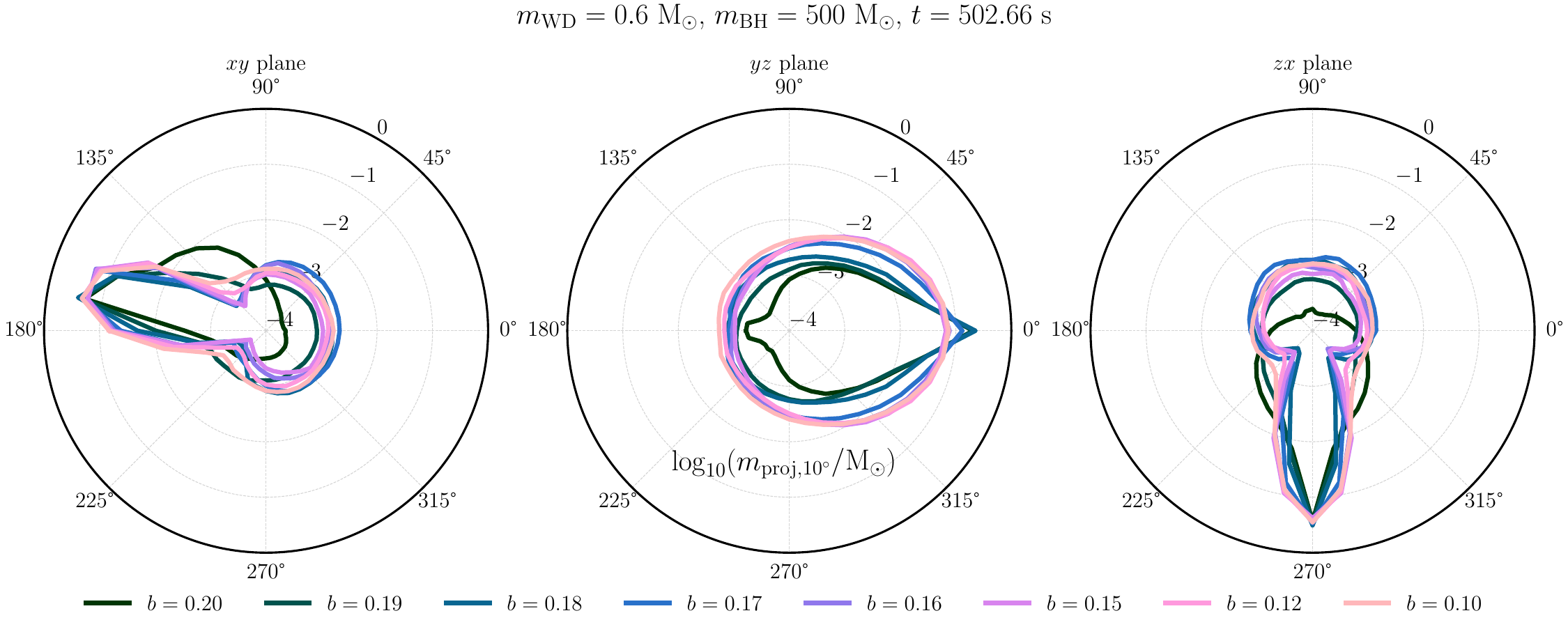}
    \caption{Polar plots of $\mproj$ similar to Figure \ref{fig:mass_asym_0.15} with the NRN enabled for different TDE $b$ values. It is clear that the ejected material is asymmetrically distributed in all cases. As $b$ decreases, the ejecta spread out more due to greater energy injection from nuclear fusion. This change is most drastic between $b = 0.20$ (standard TDE with negligible nuclear burning) and $b = 0.19$ (some nuclear burning).}
    \label{fig:mass_asym}
\end{figure*}

\begin{figure*}
    \centering
    \includegraphics[width=\textwidth]{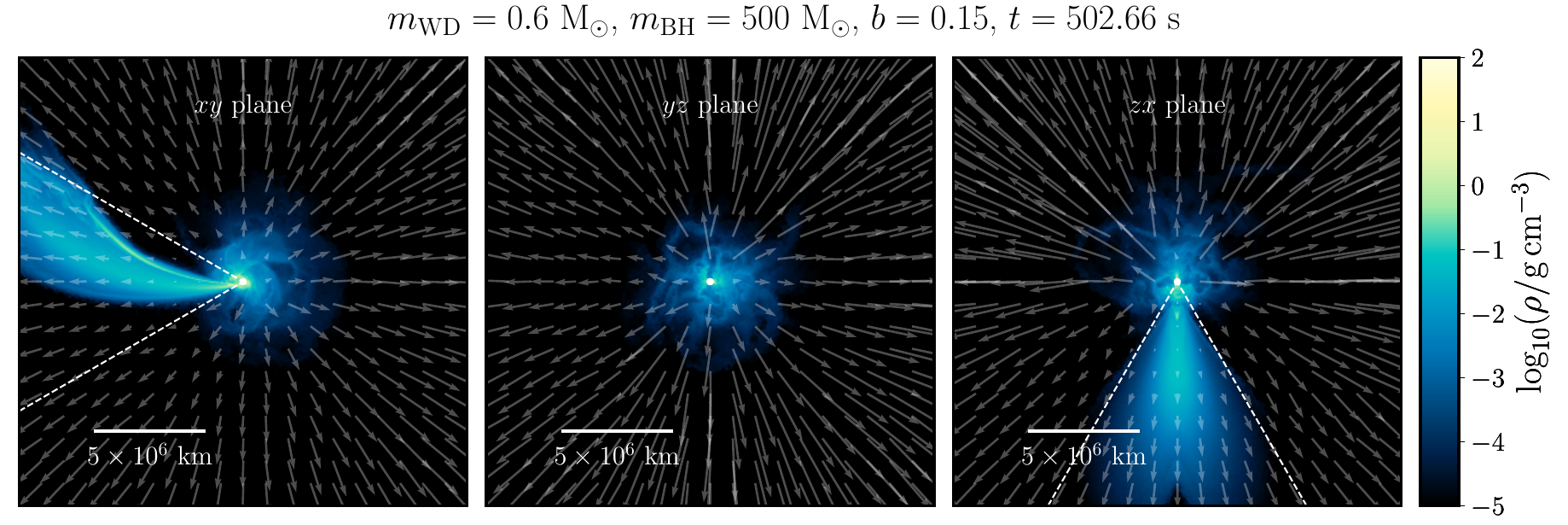}
    \caption{Snapshots of slices of density of a WD TDE with $b = 0.15$ at $\sim 500 \s$. The three panels represent slices, passing through the IMBH, in the $xy$ (left), $yz$ (center), and $zx$ (right) planes. The arrows represent the plane-projected radial velocities of the debris with respect to the IMBH at the plot centers. An arrow as long as the panel side length represents the speed of light. The dashed lines in the left and right columns represent a cone of $30^{\circ}$ oriented along the $-x$ direction -- this cone contains almost all of the debris from the TDE and the subsequent SN.}
    \label{fig:collage_vel_b0p15}
\end{figure*}

\subsection{Ejecta geometry and dynamics}
A direct consequence of highly exothermic nuclear fusion reactions is the injection of internal energy into the ejecta. The ejecta spread outward and expand ballistically, as in an SN explosion. However, due to the inherent asymmetry of the WD-IMBH encounter, the geometry of the ejecta is also highly asymmetric.

Figure \ref{fig:mass_asym_0.15} shows polar plots of plane-projected masses within wedges of opening angle $10^{\circ}$, $\mproj$, unbound from the IMBHs in three perpendicular planes $xy$, $yz$, and $zx$ at $\sim 500 \s$, for $b = 0.15$ with and without enabling the NRN. Changing the opening angle changes the absolute values of $\munb$, but not the relative differences. As a reference, the encounters take place on the $xy$ plane with the IMBH at the center; the WD arrives from the ($-x$,$-y$) direction, periapsis approach is in the $+x$ direction (very close to the IMBH), and the material is ejected mostly in the ($-x$,$+y$) direction. When the NRN is disabled, no nuclear reactions occur and, hence, a standard TDE, displaying tidal streams directed towards and away from the IMBH, ensues. Thus, the mass is directed outward in a thin tidal stream. On the other hand, when the NRN is enabled, the ejecta spread out more in all directions, including the $+z$ and $-z$ directions. However, the mass distribution is still highly asymmetric, with most of the mass being ejected in a cone of $30^{\circ}$ about the $-x$ direction, angled towards the $+y$ direction.

Figure \ref{fig:mass_asym} shows similar polar plots for different values of $b$. The ejecta geometry when $b = 0.20$ is similar to when the NRN is disabled (Figure \ref{fig:mass_asym_0.15}). As $b$ decreases, it can be seen that the ejecta fan out not only in the $+z$ and $-z$ directions, but also in the $xy$ plane.

The ejecta expand outward from the IMBH with very high velocities of $\sim 0.05\,c$, where $c$ is the speed of light. Figure \ref{fig:collage_vel_b0p15} is a collage of slices, passing through the IMBH, of the $b = 0.15$ encounter in the $xy$, $yz$, and $zx$ planes at $\sim 500 \s$. The arrows in each panel represent the projected radial velocities of the ejecta with respect to the IMBH in the respective planes. Almost all of the ejecta, except those bound to the IMBH, move outward ballistically. As inferred from Figures \ref{fig:mass_asym_0.15} and \ref{fig:mass_asym}, most of the ejecta are contained within a cone of $30^{\circ}$ about the $-x$ direction. In the following sections, we focus mainly on this region.

In a standard TDE where the initial orbit of the star is assumed to be parabolic, $\sim 50\%$ of the initial mass is bound to the BH, and $\sim 50\%$ becomes unbound \citep[e.g.,][]{1988Natur.331..687H,1988Natur.333..523R}. However, extra energy from nuclear fusion results in larger fractions of mass becoming unbound and escaping with large velocities. Table \ref{tab:mass_energy} quantifies the dynamics of the unbound ejecta: the masses $\munb$, mass fractions $\funb$, kinetic energies $\EKunb$, and nickel masses $m$\,(\nifs). When $b = 0.20$, $\sim 56\%$ of the initial WD becomes unbound from the IMBH, similar to a standard TDE. As $b$ decreases, the unbound mass increases, with the $b = 0.10$ encounter resulting in $\sim 85\%$ of the initial WD mass becoming unbound. Therefore, the kinetic energies of the ejecta also increase with decreasing $b$, from $\sim 10^{51} \erg$ when $b = 0.20$ to $\sim 1.6\times10^{51} \erg$ when $b = 0.10$. The presented quantities are calculated at $150$ -- $200 \s$ after periapsis. At later times ($t \gtrsim 300 \s$), $\EKunb$ notably increase in the cases with little to no nuclear fusion and most fallback of material, i.e., $b = 0.20$ and $b = 0.19$, due to the formation of an evacuated region with very high velocities\footnote{Since we adopt Newtonian gravity, these velocities are likely overestimated. In reality, they would be limited by general relativity.} close to the IMBH (clearly seen in the top-left panel of Figure \ref{fig:collage_quants}, the $b = 0.20$ encounter). These high-velocity outflows are far less prominent in lower-$b$ encounters, where nuclear fusion causes the ejecta to expand and re-pressurize any evacuated region, resulting in little change to their $\EKunb$. This phenomenon is illustrated in Figure \ref{fig:kinE_unbound}, which shows the kinetic energy contribution from unbound material in different density ranges at $\sim 200 \s$ and $\sim 500 \s$, for different values of $b$. The evolution of the ejecta up to $1 \dy$ using 1D radiation hydrodynamics (see Section \ref{sect_rt}) shows fallback of some of the unbound material to the IMBH. Finally, the nickel masses increase with decreasing $b$, with almost $0.5 \Msun$ of \nifs\ being ejected when $b = 0.10$.

\begin{table*}[h!]
    \caption{Masses $\munb$, mass fractions $\funb$, kinetic energies $\EKunb$, and nickel masses $m$\,(\nifs) of the unbound ejecta of WD TDEs for different $b$ values ($\mwd = 0.6 \Msun$).}
    \label{tab:mass_energy}
    \centering
    \begin{tabular}{ccccccccc}
    \hline\hline
    Property & $b = 0.20$ & $b = 0.19$ & $b = 0.18$ & $b = 0.17$ & $b = 0.16$ & $b = 0.15$ & $b = 0.12$ & $b = 0.10$ \TBstrut \\
    \hline
       $\munb$ [$\Msun$] & $0.28$ & $0.31$ & $0.38$ & $0.43$ & $0.44$ & $0.45$ & $0.48$ & $0.47$ \Tstrut \\
       $\funb$ & $0.47$ & $0.52$ & $0.63$ & $0.72$ & $0.73$ & $0.75$ & $0.80$ & $0.78$ \\
       $\EKunb$ [$10^{51}$ erg] & $0.36$ & $0.43$ & $0.63$ & $0.88$ & $1.01$ & $1.08$ & $1.20$ & $1.37$ \\
       $m$\,(\nifs) [$\Msun$] & $0.00$ & $0.01$ & $0.05$ & $0.14$ & $0.26$ & $0.35$ & $0.47$ & $0.49$ \Bstrut \\
    \hline
    \end{tabular}
    \tablefoot{The presented values are averages of $\sim 50$ -- $100 \s$ after periapsis passage. $\EKunb$ is not constant but gradually increases during the course of the simulation.}
\end{table*}

%%%%%%%%%%%%%%%%%%%%%%%%%%%%%%%%%%%%%%%%%%%%%%%%%%%%%%
%%%%%%%%%%%%%%%%%%%%%%%%%%%%%%%%%%%%%%%%%%%%%%%%%%%%%%
%%%%%%%%%%%%%%%%%%%%%%%%%%%%%%%%%%%%%%%%%%%%%%%%%%%%%%

\section{Radiative transfer calculations for SNe Ia from TDEs}
\label{sect_rt}

This section describes an exploration for the radiative-transfer modeling of the debris obtained in the \arepo\ simulations described above. The task is challenging because of the highly asymmetric distribution of the dense, \nifs-rich material unbound in these TDEs. We present a simplified approach that captures some of the salient features of such TDEs but misses a few (detailed below). Nonetheless, it complements previous work on the radiative-transfer modeling of such asymmetric debris that was presented in \citet{2016ApJ...819....3M}. More specifically, we covered a wider range of configurations, had a more detailed ejecta composition (thanks to a more complete NRN), extended the simulations to the nebular phase ($\sim 100 \dy$), and handled nonlocal-thermodynamic equilibrium more accurately. However, our treatment of asymmetry was approximate and simplistic because we did not tackle the full 3D problem, as can be done with a Monte Carlo approach. Instead, we used a combination of 1D and 2D, grid-based codes for the radiative transfer to capture the strong asymmetry of WD TDEs and predict the associated observables.
  
As shown in Figure \ref{fig:collage_quants}, there is a density contrast of order 10$^4$ or more between the material ejected along the direction $-x$ and the material that surrounds it.  Our approach is to approximate this complex geometry of the ejecta (this elongated plume) as being 2D and axisymmetric, with the bulk of the mass ejected along a cone with its tip at the center of the BH. We decided to model the radiative transfer for this 2D configuration using the 2D, steady-state polarized radiative-transfer code \longpol\ \citep{1994A&A...289..492H,1996A&A...308..521H}, upgraded to model supernova (SN) ejecta \citep{2011MNRAS.415.3497D} and applied to Type II-Plateau SNe \citep{2021A&A...651A..19D} as well as interacting, Type IIn SNe \citep{2025A&A...696L..12D}. The 2D code \longpol\ requires inputs from the 1D nonlocal thermodynamic equilibrium time-dependent radiative transfer code \cmfgen\ \citep{1998ApJ...496..407H,2012MNRAS.424..252H}, which itself requires as input an ejecta in homologous expansion. Given that there is essentially no observation of TDEs (or SNe) at very early times, time-dependent \cmfgen\ calculations are typically started at $1 \dy$ or later. 
  
Because the \arepo\ simulations were stopped at about $500 \s$, all debris needed to be evolved first until $1 \dy$. In this section, we describe in detail each of these preparatory steps, providing along the way a number of results concerning the dynamical evolution of the debris as well as the predictions for the light curves and optical spectra of the 1D counterpart to the debris from the \arepo\ simulations (Section~\ref{sect_cmfgen}). We then describe the methodology of the 2D radiative-transfer calculations with \longpol\ as well as the results in Section~\ref{sect_longpol}.

\begin{figure}
    \includegraphics[width=0.49\textwidth]{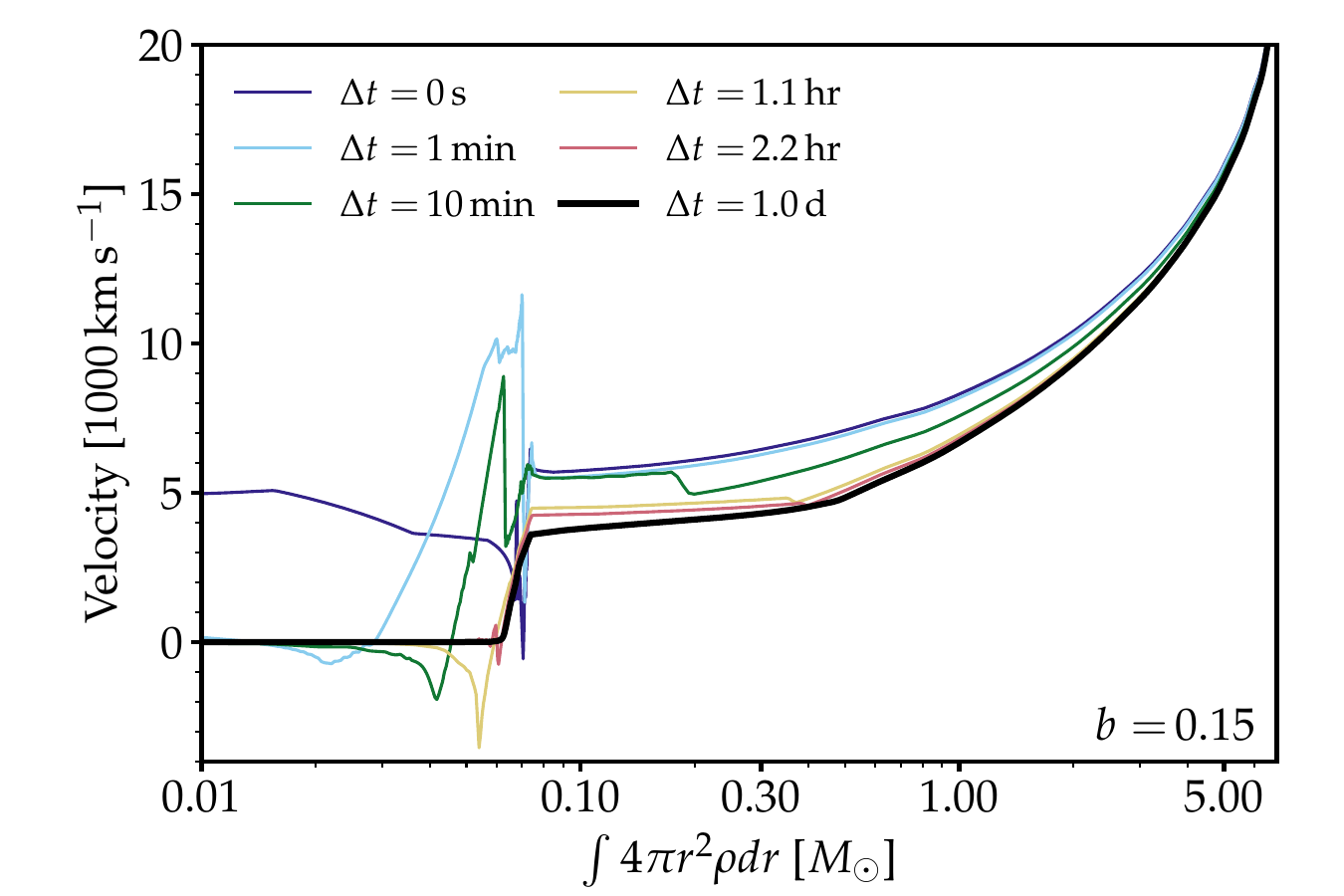}
    \includegraphics[width=0.49\textwidth]{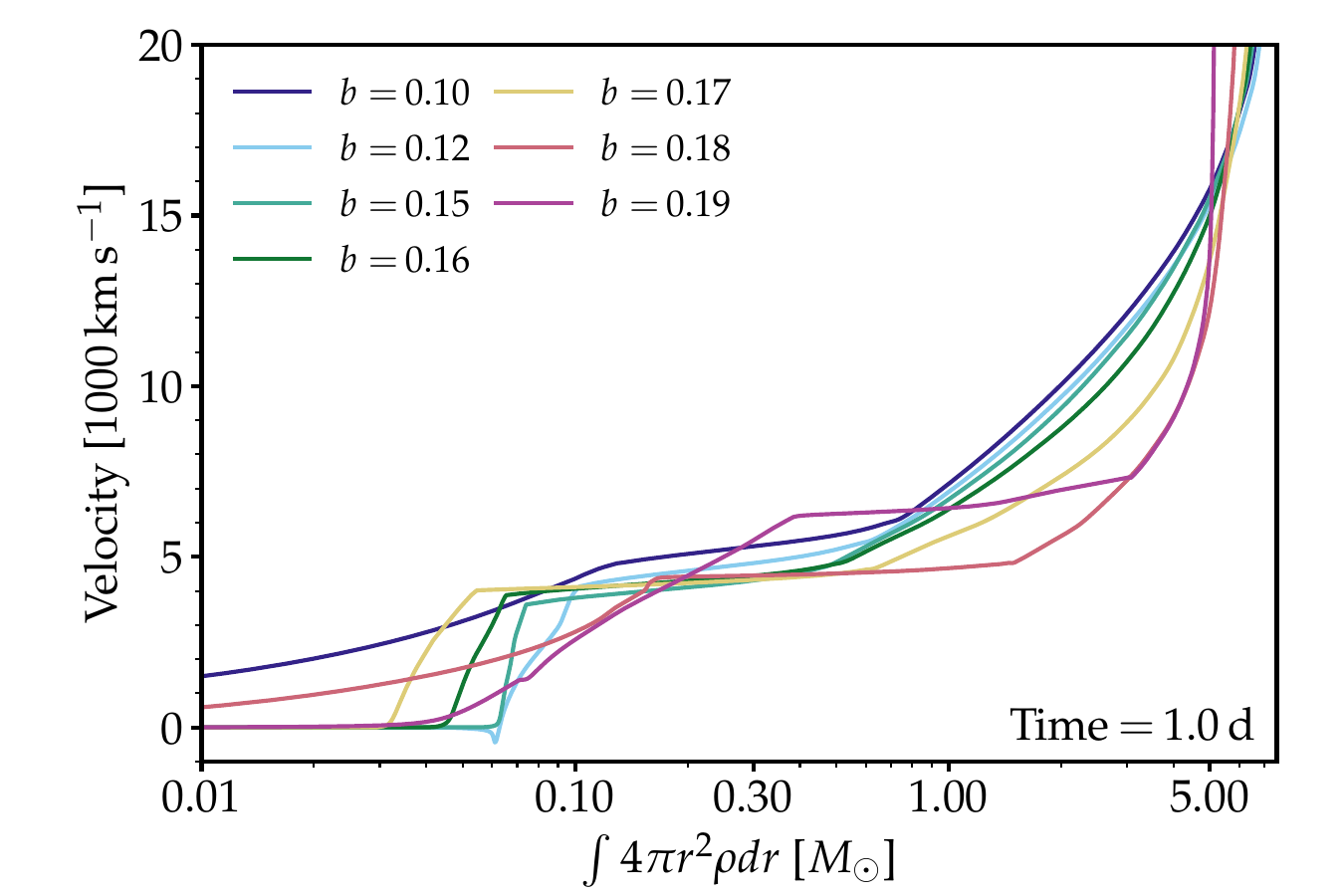}
    \caption{Velocity versus ``spherical'' mass (i.e., $\int 4 \pi r^2 \rho dr$) obtained with the radiation-hydrodynamics code \voned for the $b = 0.15$ case at multiple times up to $1 \dy$ (top) and the full model set at $1 \dy$ (bottom). A velocity hole is observed in the unbound debris below $\sim 3000 \kmsinv$ due to fallback of material.}
    \label{fig:fallback_v1d}
\end{figure}

\subsection{Preparation of inputs at $1 \dy$}

For the last snapshot at $\sim 500 \s$ of each \arepo\ simulation, we performed spherical shell-averaging around the IMBH to obtain the 1D radial profiles of density, temperature, radial velocity, and isotopic abundances of the ejecta (hence all as a function of distance from the IMBH). Only the cells unbound from the IMBH (i.e., those with a positive total energy, which sums the kinetic, potential, and internal energy parts) were included in the calculation. In addition, since almost all ejecta are in a cone of $30^{\circ}$ oriented along the $-x$ direction (see Figure \ref{fig:collage_vel_b0p15}), we only calculated the shell-averages for cells within the cone.

The angular average of the distribution of material along the $-x$ direction at $\sim 500 \s$ was remapped into the radiation-hydrodynamics code \voned\ \citep{1993ApJ...412..634L,2010MNRAS.405.2113D}. In the process, we averaged the composition within each radial slice, hence losing the heterogeneous composition present in the 3D debris. As shown in Figure \ref{fig:collage_nuc}, unburnt material (i.e., C and O) is located in a shell outside the metal-rich, internal regions of the asymmetric ejecta, and thus stands apart from explosive ashes traced by, for example, \nifs. However, because of the peculiar geometry, this unburnt material is present at all ejecta velocities, from the slower-moving regions close to the BH and around the plume all the way out to the largest velocities away from the BH. In contrast, unburnt material in standard, quasispherical (i.e., relative to SNe Ia TDEs) Type Ia SN ejecta survives only in the outermost ejecta layers, and thus only at the largest velocities. This different chemical stratification is a fundamental difference between the ejecta of standard SNe Ia and those from WD-TDEs.

\begin{figure*}
    \centering
    \includegraphics[width=0.49\textwidth]{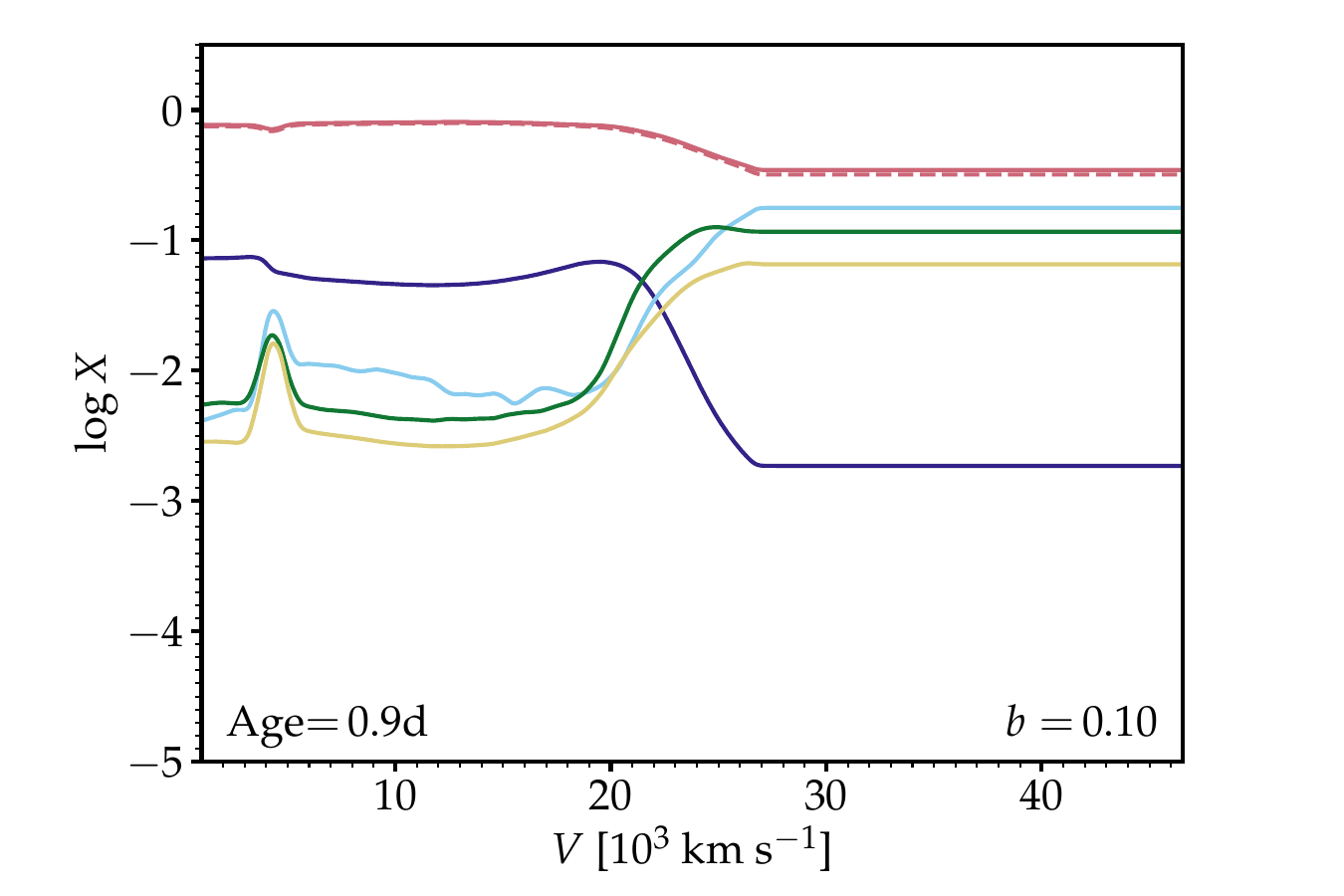}
    \includegraphics[width=0.49\textwidth]{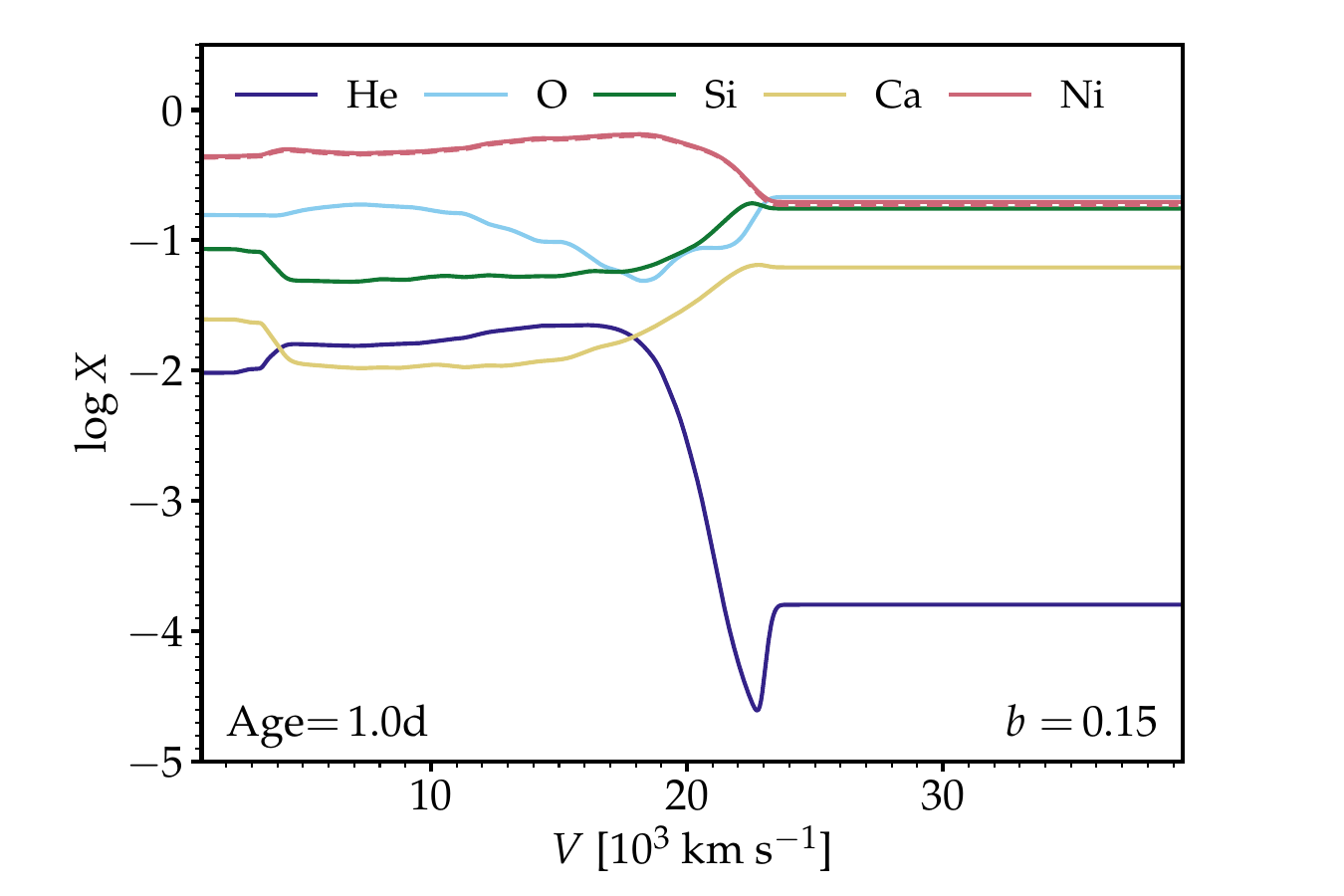}
    \vspace{0.01cm}
    \includegraphics[width=0.49\textwidth]{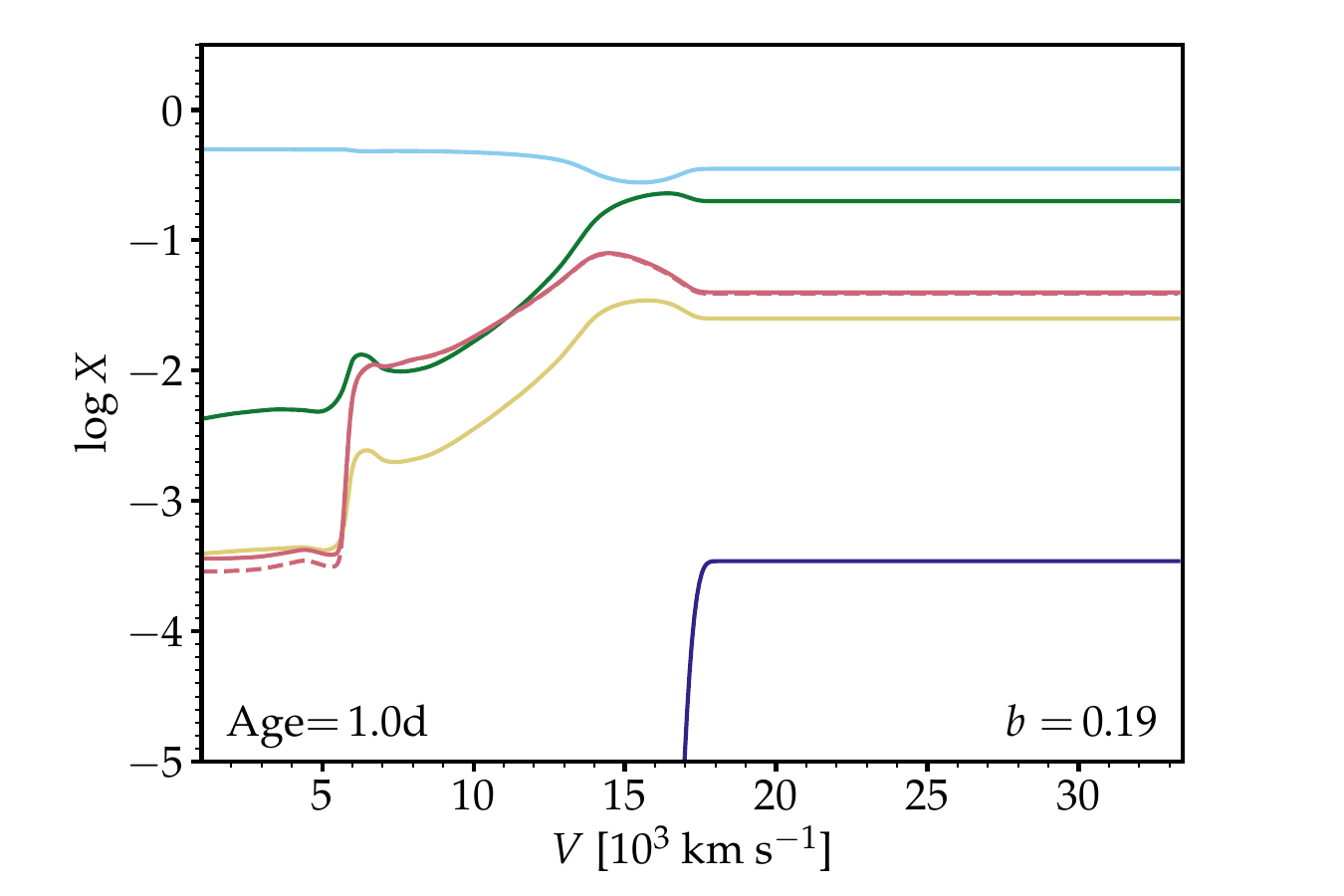}
    \includegraphics[width=0.49\textwidth]{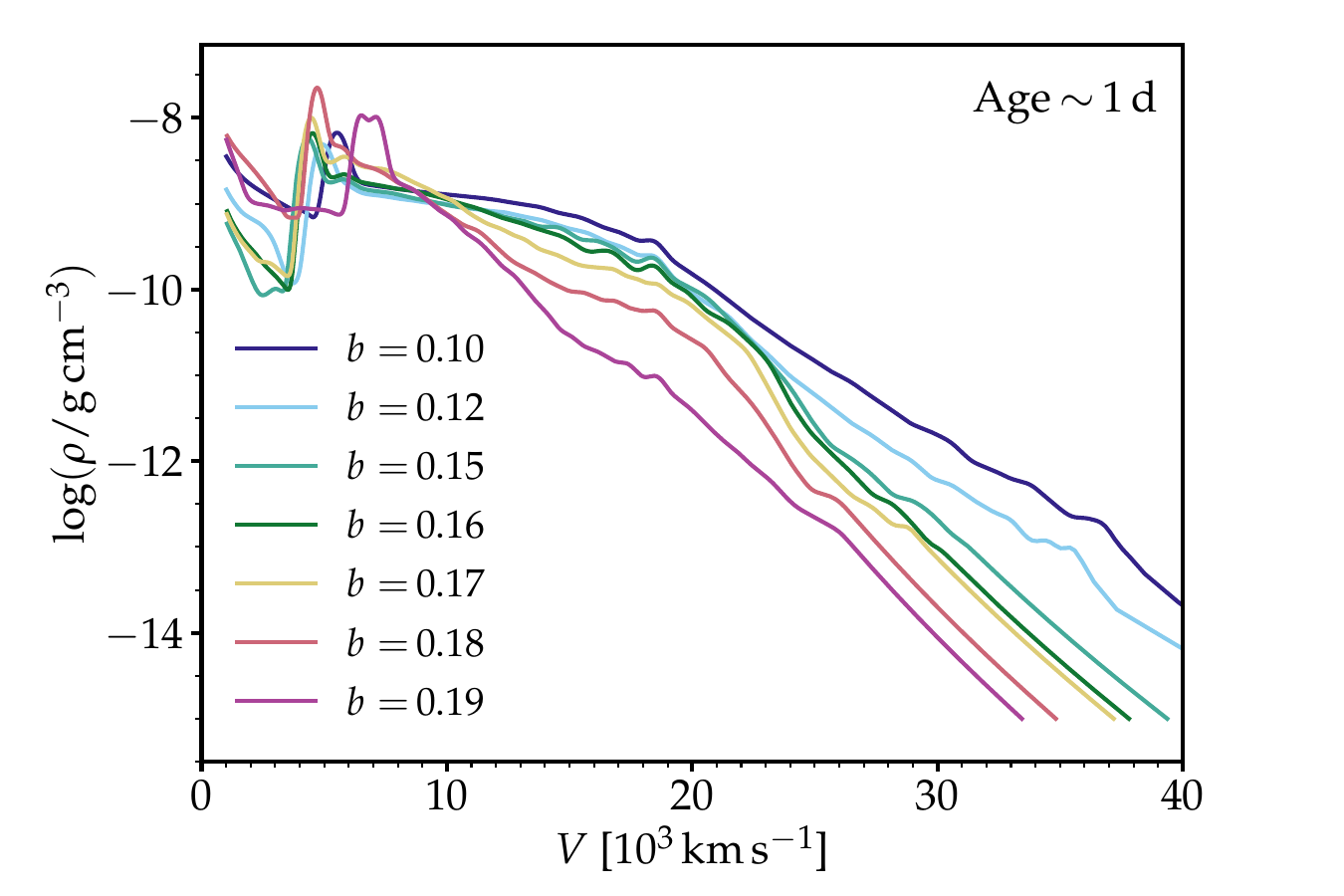}
    \caption{Composition profiles for models $b = 0.10$ (top left), $b = 0.15$ (top right), and $b = 0.19$ (bottom left), as well as density profiles for the whole model set (bottom right), all at $\sim 1 \dy$ and used as initial conditions for the 1D radiative transfer calculations. We show a few of the dominant species in the ejecta. The Ni profile accounts for radioactive decay (essentially all Ni is due to \nifs). All elements are present at different radial velocities of the ejecta.}
    \label{fig:cmfgen_init_comp}
\end{figure*}

These spherically-averaged debris were then evolved until about $1 \dy$. The $b = 0.20$ case was excluded from the sample because it contains no \nifs. This evolution was computed in the same manner as for core-collapse SN ejecta, that is, by including gravity (here from a central $500 \Msun$ BH rather than a $\sim 1.5 \Msun$ neutron star) and by allowing for radioactive decay from unstable nuclei (primarily from \nifs). We ignored any power injection from the BH (e.g., as from fallback accretion) -- the BH was assumed to be dormant throughout. This preparatory step also neglects any multi-dimensional effect -- the material along the considered direction behaves as if its spatial distribution were spherically symmetric. To attenuate the large composition gradients in the debris (in particular the sharp jumps that develop in the inner ejecta), we smoothed the composition at $1000 \s$ with a boxcar width of $0.05 \Msun$. Because of the assumption of spherical symmetry, all ejecta masses (total and isotopic) are typically 10-20 times larger than in the original 3D \arepo\ simulation counterpart. 

In contrast to Type Ia SN ejecta, which reach homologous expansion after a few tens of seconds, a SN Ia TDE is influenced by the gravitational pull of the central (here of $500 \Msun$) BH, and the ejecta are still far from reaching homologous expansion at the end of our simulations. Consequently, even though the input from \arepo\ was truncated at small radii to include only unbound material at $\sim 500 \s$, we found that some material falls back into the BH during the subsequent evolution. As illustrated for the $b = 0.15$ case in the top panel of Figure \ref{fig:fallback_v1d}, there are two visible effects. First, about a tenth of a solar mass falls back within a few hours after the start of the simulation, which creates a central ``hole'' in the debris since no material is ejected with velocities below about $3000 \kmsinv$. Second, the velocities of the debris layers above drop by up to a factor of two during this expansion out of the gravitational potential well, reaching homologous expansion ($V \propto r$) only after several hours. Further out in the debris, this pull is weaker, and the material follows a ballistic trajectory earlier. The inner velocity hole varies in size within our model set (bottom panel of Figure \ref{fig:fallback_v1d}). It is the largest in the model with the lowest \nifs\ mass, which also has the lowest kinetic energy and the highest density in the inner debris at $500 \s$. In other models, the jump varies in size and extent, with no clear monotonic relation to \nifs\ mass. 

Physically, this fallback is related to the large escape speed of a few $1000 \kmsinv$ for a $500 \Msun$ BH even at large radii of $10^{7}$ -- $10^{8} \km$. Observationally, it is an important signature since it not only connects to the BH mass but would also have a spectroscopic signature since no spectral line emission or absorption can arise from these essentially void inner regions.

\subsection{Results for the 1D-equivalent models computed with \cmfgen}
\label{sect_cmfgen}

\begin{figure*}
    \centering
    \includegraphics[width=0.49\textwidth]{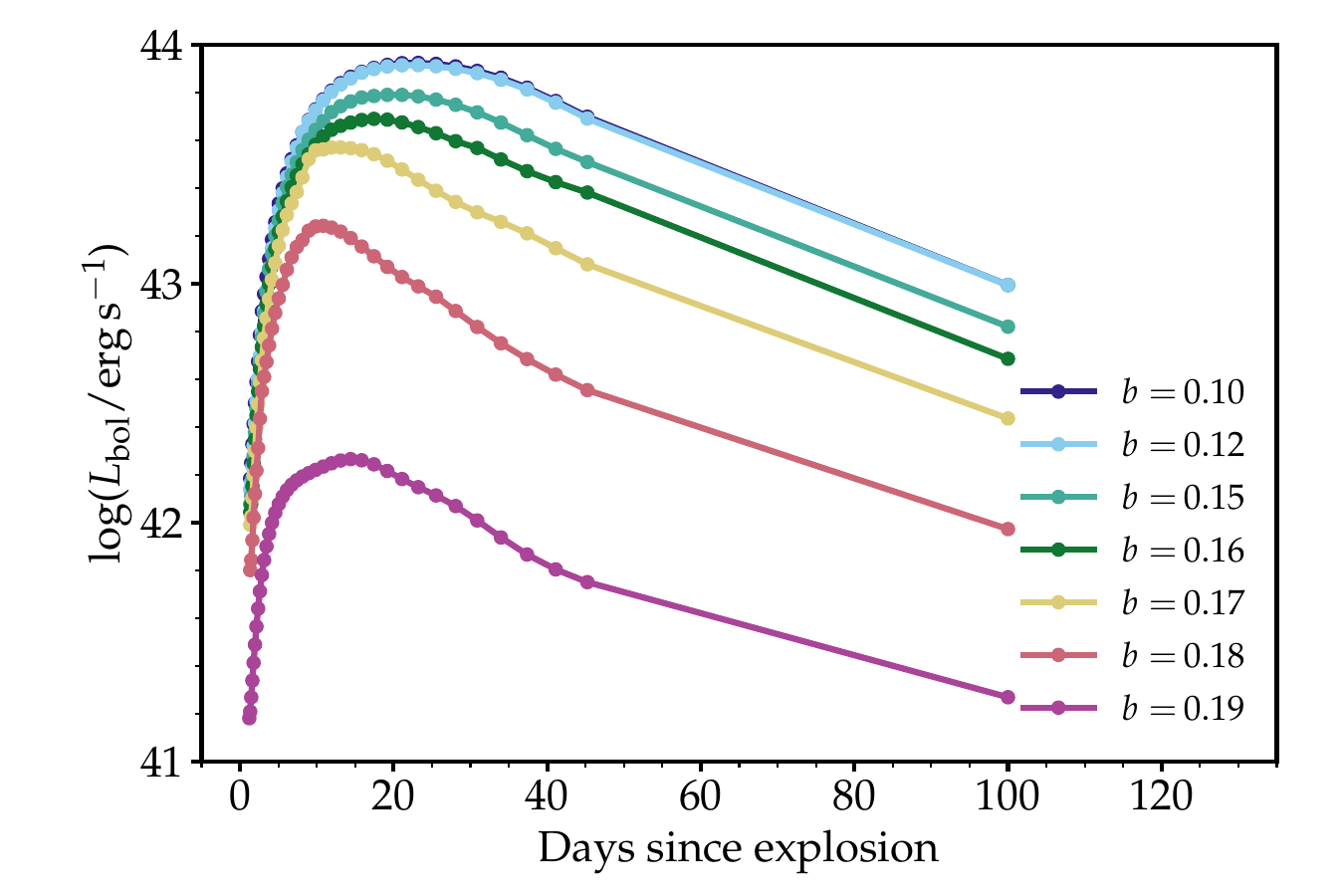}
    \includegraphics[width=0.49\textwidth]{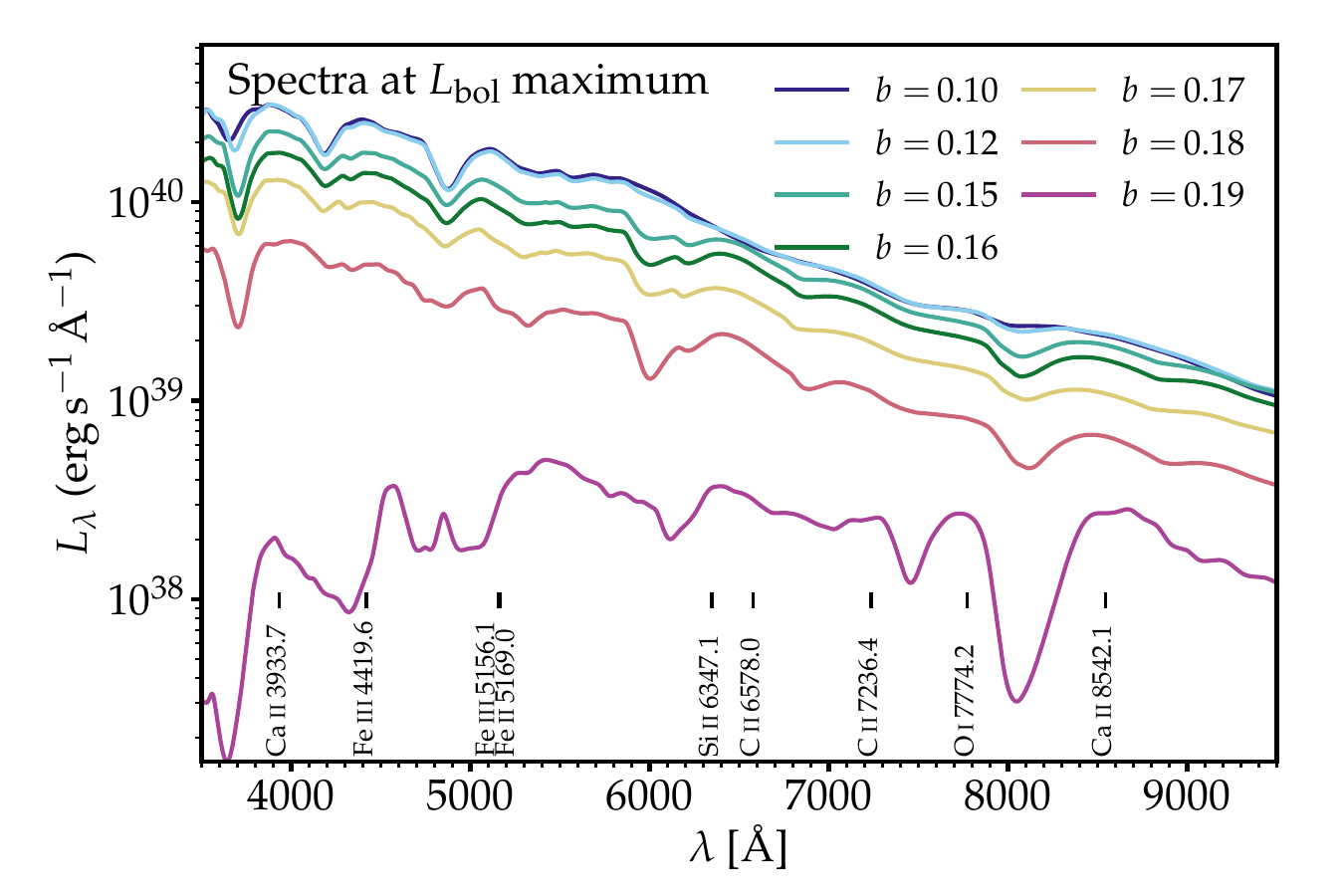}
    \caption{Radiative properties of the 1D counterpart calculated with \cmfgen\ and based on the 3D models computed with \arepo. We show the bolometric light curves (left) and the maximum-light optical spectra for our model set (right). Closer encounters (e.g., $b = 0.10$) are brighter and show stronger Fe\three\ lines in their spectra, while wider encounters (e.g., $b = 0.19$) are dimmer and show stronger O\one, Si\two, and Ca\two\ lines in their spectra.}
    \label{fig:lbol_spec_at_max}
\end{figure*}

\begin{figure*}
    \sidecaption
    \includegraphics[width=0.7\textwidth]{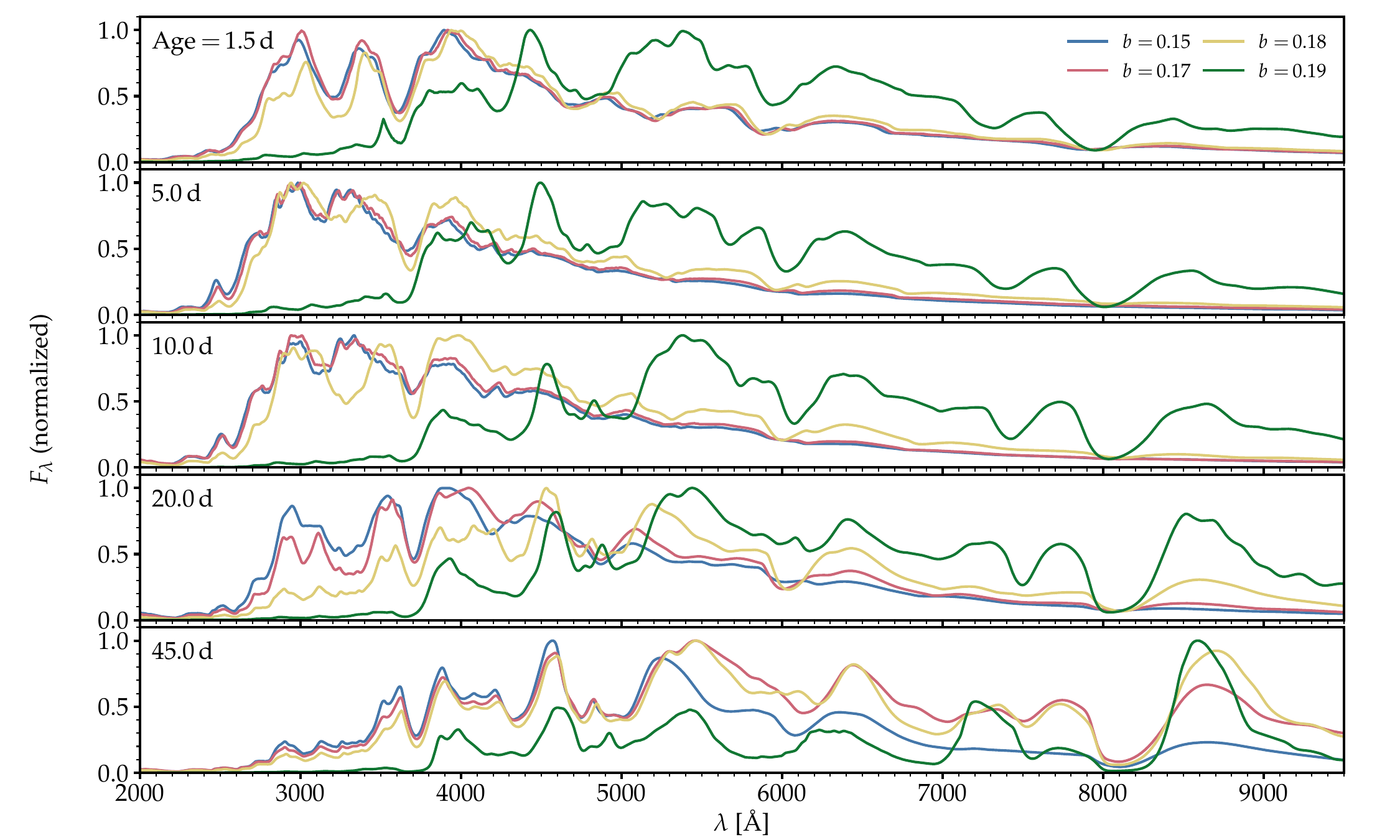}
    \caption{Montage of spectra at $1.5$, $5$, $10$, $20$, and $45 \dy$ for the 1D models when $b = 0.15$, $0.17$, $0.18$, and $0.19$. These epochs cover the high-brightness optically thick phase. For line identifications, see the other figures.}
    \label{fig:spec_all}
\end{figure*}

\begin{figure}
    \includegraphics[width=0.49\textwidth]{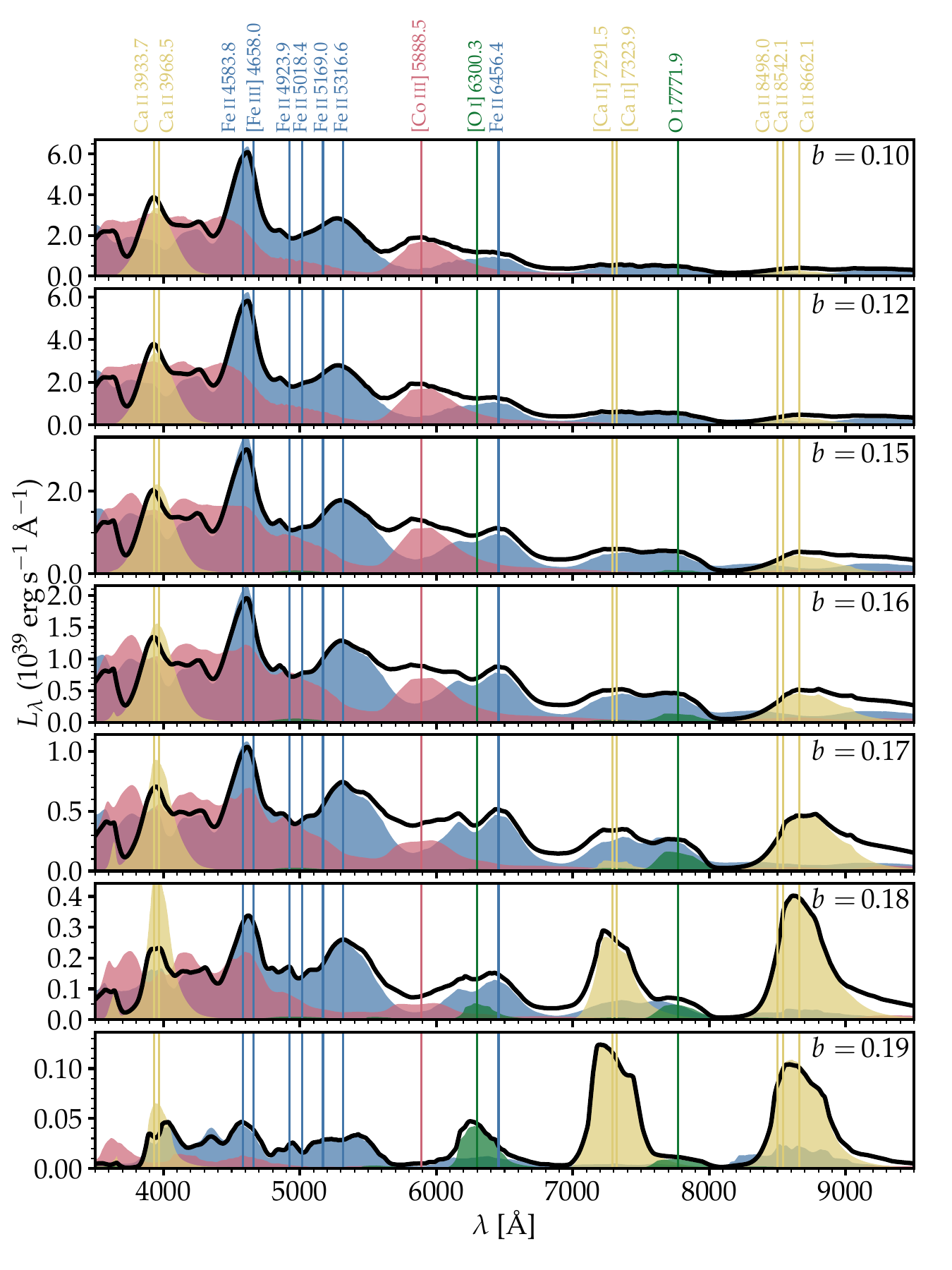}
    \caption{Montage of spectra for the 1D-equivalent of the 3D models computed with \cmfgen\ at $100 \dy$ after the tidal disruption. A color coding is used to differentiate the emission from O, Ca, Co, and Fe, which dominate at that late time. The spectra of closer encounters with larger \nifs\ masses (e.g., $b = 0.10$) are dominated by Co\three\ and Fe\three\ lines, while those of wider encounters with lower \nifs\ masses (e.g., $b = 0.19)$ are dominated by O\one\ and Ca\two\ lines.}
    \label{fig:spec_100d}
\end{figure}

The results from \voned\ were mapped into \cmfgen\ about one day after the disruption and explosion of the white dwarf.  We proceeded using the standard technique, as described, for example, for similar simulations of SNe Ia in \citet{2023A&A...678A.170B}, in particular, for the updated atomic data and model atoms that we adopted verbatim from that work. Homologous expansion was strictly enforced. We accounted for important species, including He, C, N, O, Ne, Na, Mg, Al, Si, S, Ar, Ca, Sc, Ti, V, Cr, Mn, Fe, Co, and Ni. We allowed only for the decay of \nifs\ and \cofs\ since \nifs\ vastly dominates in both decay power and abundance compared to any other unstable isotope produced in those explosions and for the times considered here.

Figure \ref{fig:cmfgen_init_comp} shows the initial composition profile for He, O, Si, Ca, and Ni at the start of the \cmfgen\ simulation at $1 \dy$ for models $b = 0.10$ (i.e., the model with the highest \nifs\ mass), $b = 0.15$ (i.e., an intermediate case), and $b = 0.19$ (i.e., the model with the lowest, though nonzero \nifs\ mass), as well as the density profile for the full set of models. The central cavity, bounded from above by a narrow dense shell up to 100 times denser, is visible in the density profiles. The composition profiles show little chemical stratification (this effect is artificially enhanced by the adopted spherical averaging) so that all elements are typically present with at least a mass fraction of 1\% at any location in the ejecta -- this arises from the peculiar chemical segregation in SNe Ia TDEs (see earlier discussion and Figure \ref{fig:collage_nuc}). This will matter when analyzing the spectral information since a given element may emit or absorb radiation and thus contribute to the local cooling at any location in the debris. 

Because of the 1D modeling of the highly-asymmetric 3D debris, the (spherically-extrapolated) ejecta energy and masses in the 1D \cmfgen\ models are much greater than typically encountered in SNe Ia. The 1D model corresponding to $b = 0.10$ has an ejecta mass of $7.19 \Msun$, a kinetic energy of $1.41 \times 10^{52} \erg$, and a \nifs\ mass of $6.11 \Msun$. At the opposite end, the $b = 0.19$ model has a similar ejecta mass of $5.16 \Msun$, a lower kinetic energy of $3.6 \times 10^{51} \erg$, and a much lower \nifs\ mass of $0.085 \Msun$ -- this spherical mass is typical of core-collapse SNe and should produce a peak luminosity comparable to what is observed in SNe Ibc. In contrast, the $b = 0.10$ model  should produce a large luminosity.

The left panel of Figure \ref{fig:lbol_spec_at_max} shows the bolometric light curves for the \cmfgen\ simulations of the 1D counterpart of the 3D \arepo\ simulations for the $b = 0.10$, $0.12$, $0.15$, $0.16$, $0.17$, $0.18$, and $0.19$ models (model $b = 0.20$ was not considered because it is \nifs\ free). Our model set covers rise times of $\sim 14 \dy$ to $\sim 23 \dy$ to a bolometric maximum of $1.8 \times 10^{42} \ergsinv$ to $8.3 \times 10^{43} \ergsinv$. All simulations up to the last time of $45.2 \dy$ were computed with the time-dependent solver in \cmfgen. An additional epoch was computed, but in steady state at $100 \dy$, where the conditions are nebular. At that time, the light curve shows a flattening because an increased fraction of the decay power arises from locally absorbed positrons from the decay of \cofs.  

The optical spectra at bolometric maximum (right panel of Figure \ref{fig:lbol_spec_at_max} are naturally ordered in decreasing flux from $b = 0.10$ to $b = 0.19$, all showing broad P-Cygni type line profile morphologies. The $b = 0.10$ model and analogs at high \nifs\ mass exhibit a dominance of Fe\three\ lines (e.g., at $4419.6 \ang$ and $5156.1 \ang$, but there are a myriad of lines, including contributions from Co\three\ at that time) whereas the $b = 0.19$ model and analogs at low \nifs\ mass exhibit a dominance of lines from intermediate mass elements with O\one\ (e.g., at $7771.4 \ang$), Si\two\ (e.g., at $6355.2 \ang$), and Ca\two\ (e.g., the near-infrared triplet that spreads over the $8000$ -- $9000 \ang$ region). In between, our models show a mixture of both sets of lines and produce more complex line profiles, for example, with lines of Si\two\ and Fe\two\ -- Fe\three\ around $6500 \ang$.

The spectral evolution computed with \cmfgen\ for the $b = 0.15$, $0.17$, $0.18$, and $0.19$ models at $1.5$, $5$, $10$, $20$ and $45 \dy$ is shown in Figure \ref{fig:spec_all}. The $b = 0.10$ and $b = 0.12$ models with larger \nifs\ masses than the $b = 0.15$ model have a similar spectral morphology and a global offset in flux (thus not shown). Most of the diversity occurs at the lower end of the \nifs\ distribution where the contribution from IMEs strengthens. This is made more evident in Figure \ref{fig:spec_100d} where all models computed with \cmfgen\ are shown at $100 \dy$. The difference at this later time is the additional presence of forbidden lines, primarily of [Co\three]\,$5888.5 \ang$ and [Fe\three]\,$4658.0 \ang$ in models with a high \nifs\ mass whereas [O\one]\,$6316.0 \ang$ and [Ca\two]\,$7307.7 \ang$ dominate in our models with modest amounts of \nifs. Hence, this grid effectively covers from SNe Ia ejecta entirely dominated by IGEs similar to 91T-like events (for a review, see \citealt{2024ApJS..273...16P}), to standard Type Ia SN ejecta (e.g., SN\,2002bo; \citealt{2004MNRAS.348..261B}; \citealt{2015MNRAS.448.2766B}), and down to ejecta that are more similar in composition to Type Ic SNe (C and O dominated with little \nifs; e.g., \citealt{1994Natur.371..227N}).

This evolution in time and relative to \nifs\ mass is similar to what has been shown in numerous other studies for the explosion of both Chandrasekhar and sub-Chandrasekhar white dwarfs (e.g., \citealt{1996ApJ...457..500H,2010ApJ...714L..52S,2013MNRAS.429.2127B,2013ApJ...770L...8P,2014MNRAS.441..532D}) and does not warrant further discussion. It is done here to set the stage for the later discussion on the 2D simulations.

\section{Results from 2D radiative-transfer calculations}
\label{sect_longpol}

\begin{figure*}
    \includegraphics[width=0.49\textwidth]{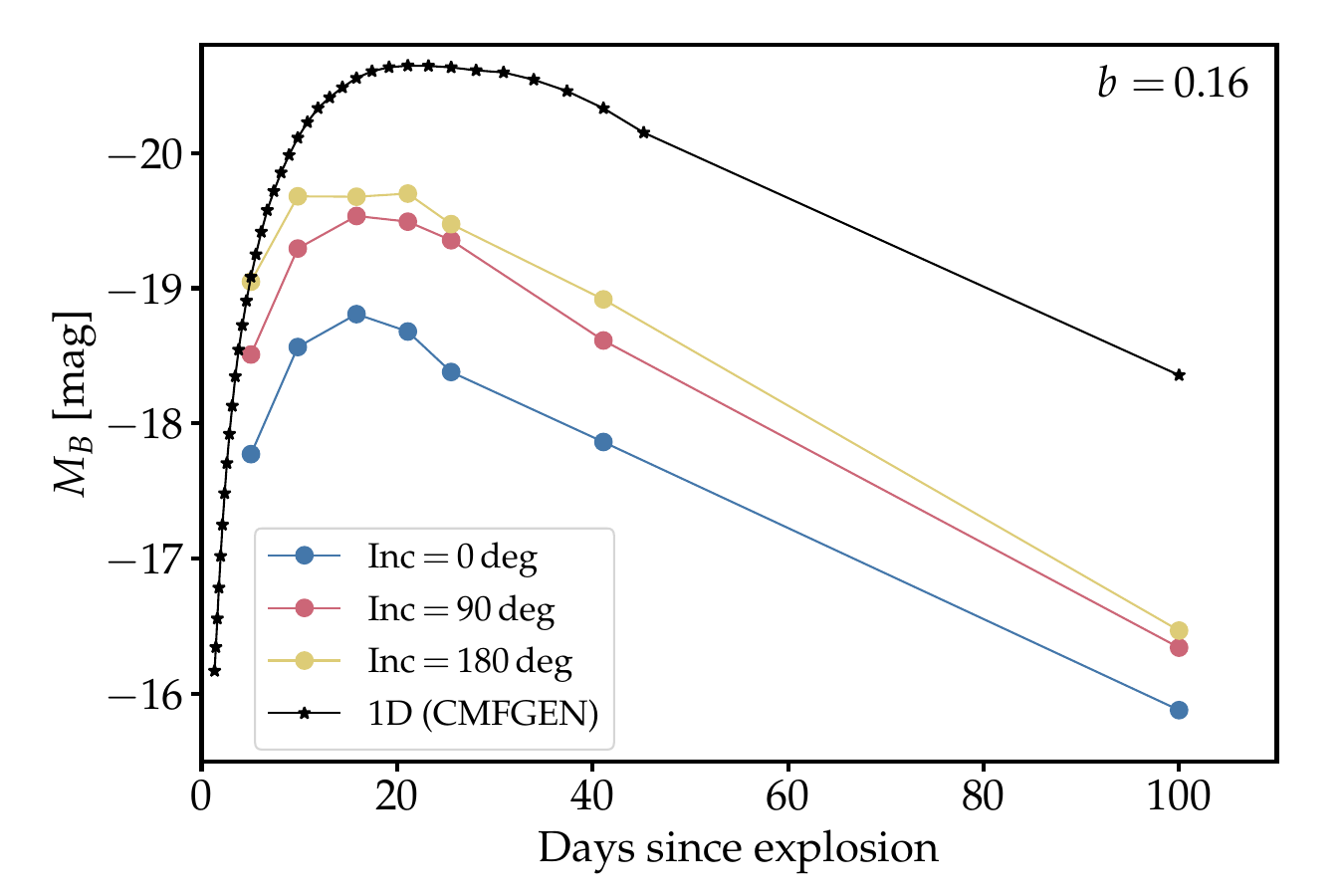}
    \includegraphics[width=0.49\textwidth]{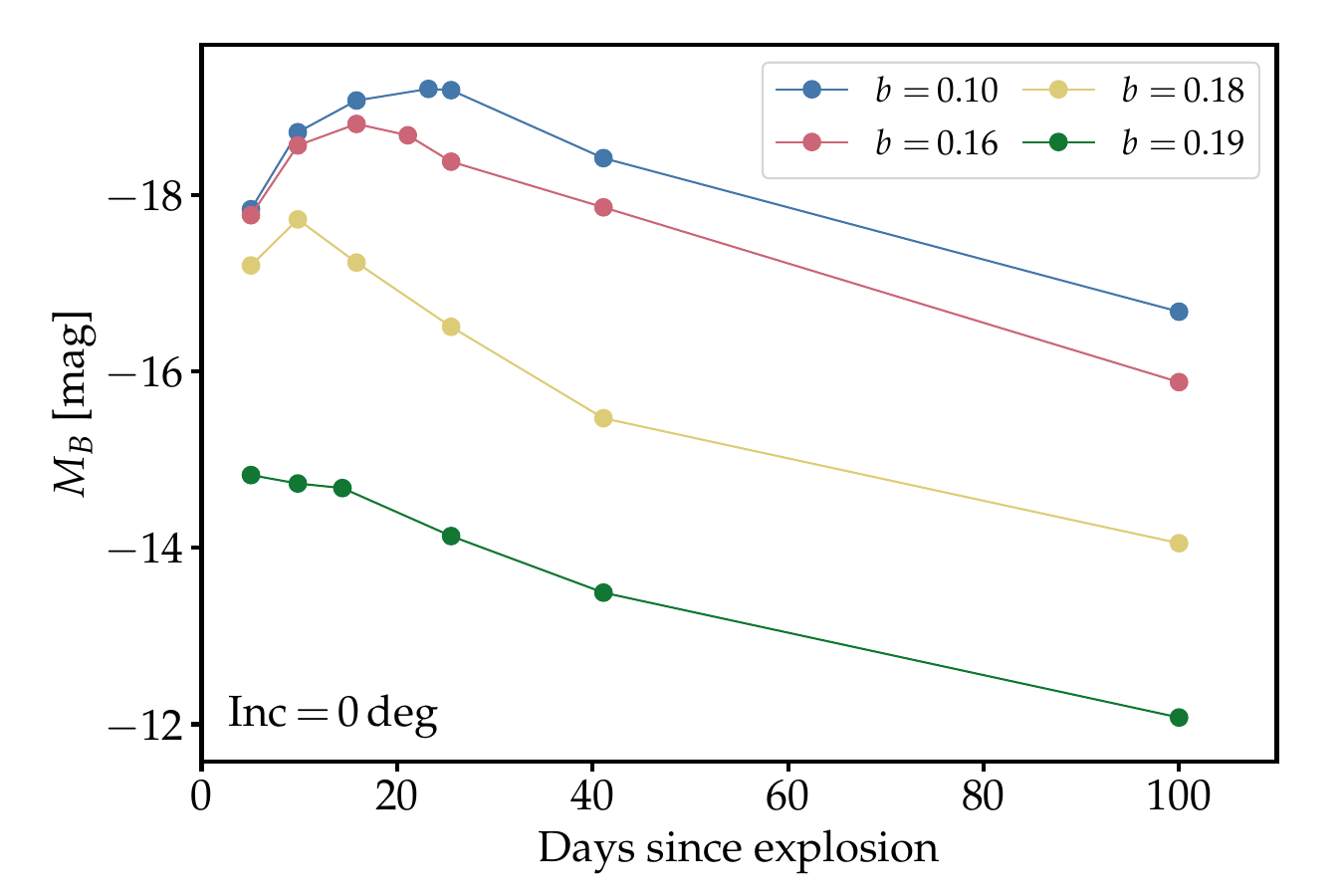}
    \includegraphics[width=0.49\textwidth]{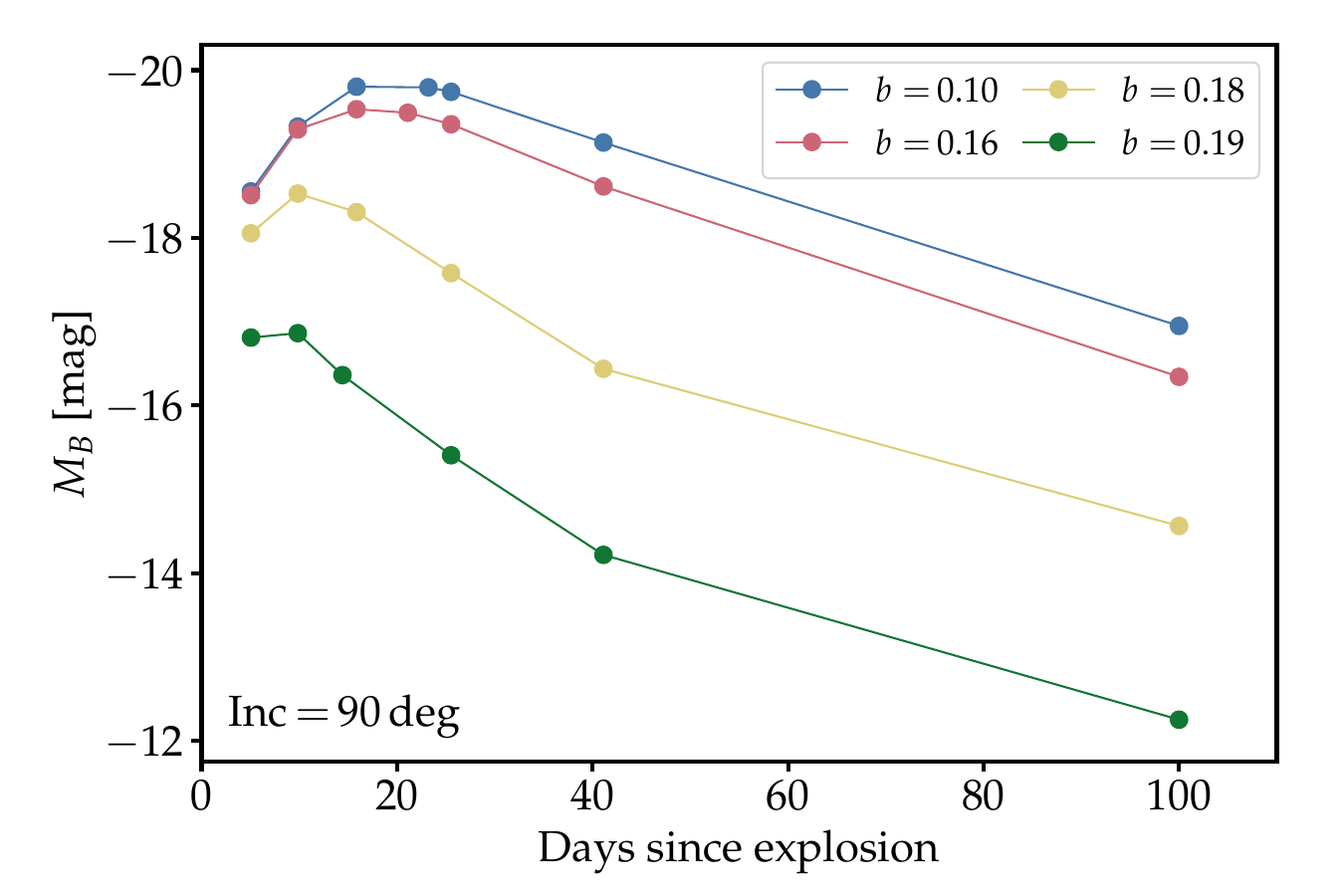}
    \includegraphics[width=0.49\textwidth]{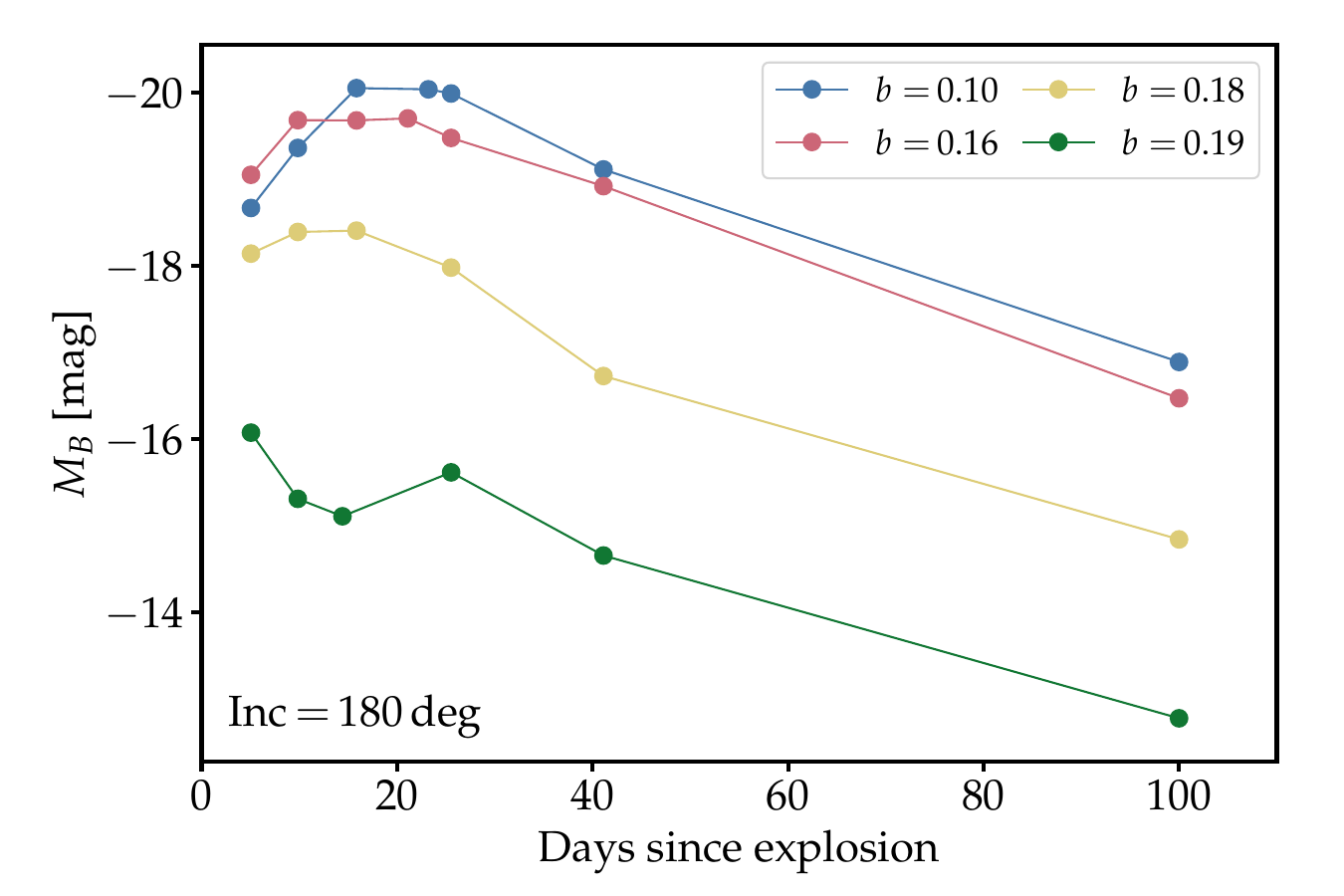}
    \caption{Photometric properties of SNe Ia TDE models computed with \longpol. Top left: $B$-band light curve of the $b = 0.16$ model for viewing angles of $0$, $90$, and $180 \deg$. We also overplot the \cmfgen\ results (assuming spherical symmetry). Other panels: Same as top left, but comparing different models for a given viewing angle. The $B$-band brightness is greater for closer encounters and viewing angles away from $0 \deg$.}
    \label{fig:lc_long_pol}
\end{figure*}

The 1D simulations presented in the preceding section were post-processed at multiple epochs with the 2D, steady-state polarized radiative-transfer code \longpol. In practice, the 2D code reads the 1D wavelength-dependent total (line and continuum) opacities $\chi$, emissivities $\eta$, and electron densities $N_e$ versus the velocities output by the \cmfgen\ calculations and remaps these onto the 2D, spherical polar grid of \longpol. The asymmetry of the 3D \arepo\ simulations was encoded in \longpol\ through a latitudinal scaling of $\chi$, $\eta$, and $N_e$. Below a polar angle of $\theta_0$ (relative to the axis of the cone at $\theta=0$ deg), this scaling factor was set to unity, whereas beyond it was set to vary as $\alpha + (1-\alpha) \exp(-\beta^2)$, where $\beta = (\theta-\theta_0)/\Delta \theta$. To reflect the asymmetry of the 3D debris, we adopted $\alpha = 10^{-8}$, $\theta_0 = 15 \deg$, and $\Delta \theta = 10 \deg$. We found that the value of $\alpha$ had to be very low in order to quench the contribution from the material surrounding the unbound ejecta, particularly at early times when the debris is still dense and optically thick. Figure \ref{fig:lat_scale} illustrates this latitudinal scaling.

Obviously, this treatment of the asymmetry is very rough. When assuming spherical symmetry, the deeper ejecta layers are not only denser, which leads to a smaller photon mean free path, but they are also enshrouded within the overlying ejecta layers, which may contribute a significant optical depth. This affects the propagation of both $\gamma$-rays (i.e., from radioactive decay emission) and low-energy photons. When one allows for asymmetry (whether truly in 3D or in the approximate morphology in 2D), the deeper layers are equally dense as in the 1D case, but they are no longer enshrouded. This holds for the sides of the ``cone'', or for the innermost layers which are effectively bare. In our post-treatment of the 1D \cmfgen\ simulations with \longpol, we ignored this aspect and thus overestimate the temperature and ionization of these deeper layers. This means that we overestimate the emissivity of the deeper layers, which affects all observables for viewing angles away from $\theta = 0 \deg$. More specifically, we overestimate the brightness (a global shift in all bands), the color (predictions will be overly blue in the optical), and the ionization (we may predict emission from twice-ionized species, whereas once-ionized species may dominate). This issue is, however, not as bad as it may seem since the inner ejecta are $10^5$ times denser than the outer, fast-moving ejecta regions. Thus, even if the inner ejecta regions are bare, the mean-free-path of both low- and high-energy photons is small, and radiation is considerably more trapped in the inner ejecta regions than in the outer ejecta regions. A more suitable approach would be to solve the full radiative transfer of low- and high-energy photons in 3D, as done in \citet{2016ApJ...819....3M} -- this is deferred to future work.

The immediate difference that results from this 2D treatment is that the ejecta and isotopic yields are now commensurate with those from the 3D simulations (i.e., the 2D ejecta volume is reduced by a factor of about 20 compared to the spherical case). In 2D, the unbound material in the $b = 0.10$ ($b = 0.19$) model now has a total mass of $0.34 \Msun$  ($0.24 \Msun$), compared to ejecta masses in the 3D \arepo\ simulations of $0.51 \Msun$ ($0.34 \Msun$) and of $7.11 \Msun$ ($5.14 \Msun$) compared to the spherical counterparts of the preceding section. Consequently, the light curves are now more comparable with those of Type Ia SNe. Since we did not aim to reproduce the 3D debris masses exactly, we applied the same latitudinal scaling ($\alpha$, $\theta_0$, and $\Delta \theta$ parameters) for all simulations.

Figure \ref{fig:lc_long_pol} shows the $B$-band light curves computed for a variety of models and viewing angles with \longpol\ (for comparison, the spherical counterparts obtained with \cmfgen\ are included). The overall offset with the 1D results is $2$ -- $3$\,mag due to the reduced \nifs\ mass. The predicted brightness is greater for viewing angles away from $\theta = 0 \deg$. This is qualitatively similar to the results of \citet{2016ApJ...819....3M}, although they obtained a larger offset of about $2$\,mag for their model with $0.13 \Msun$ of \nifs\ (equivalent to the $b = 0.17$ model here). This dependency on viewing angle results from a combination of effects. One is the change in the size of the emitting surface seen by the observer (i.e., greater for viewing angles closer to $90 \deg$). Another is the observation of hotter, inner ejecta regions for viewing angles away from $\theta = 0 \deg$. In our work, the results for the photometry are impacted by the adoption of the emissivity from the 1D \cmfgen\ models, which underestimates photon escape (and thus cooling) from the inner ejecta layers. Interestingly, the photometric offset is still present at $100 \dy$ when conditions are nebular (the total inward-integrated radial Rosseland-mean optical depth is 0.9 in the model $b = 0.16$ at $100 \dy$).

\begin{figure}
    \includegraphics[width=0.49\textwidth]{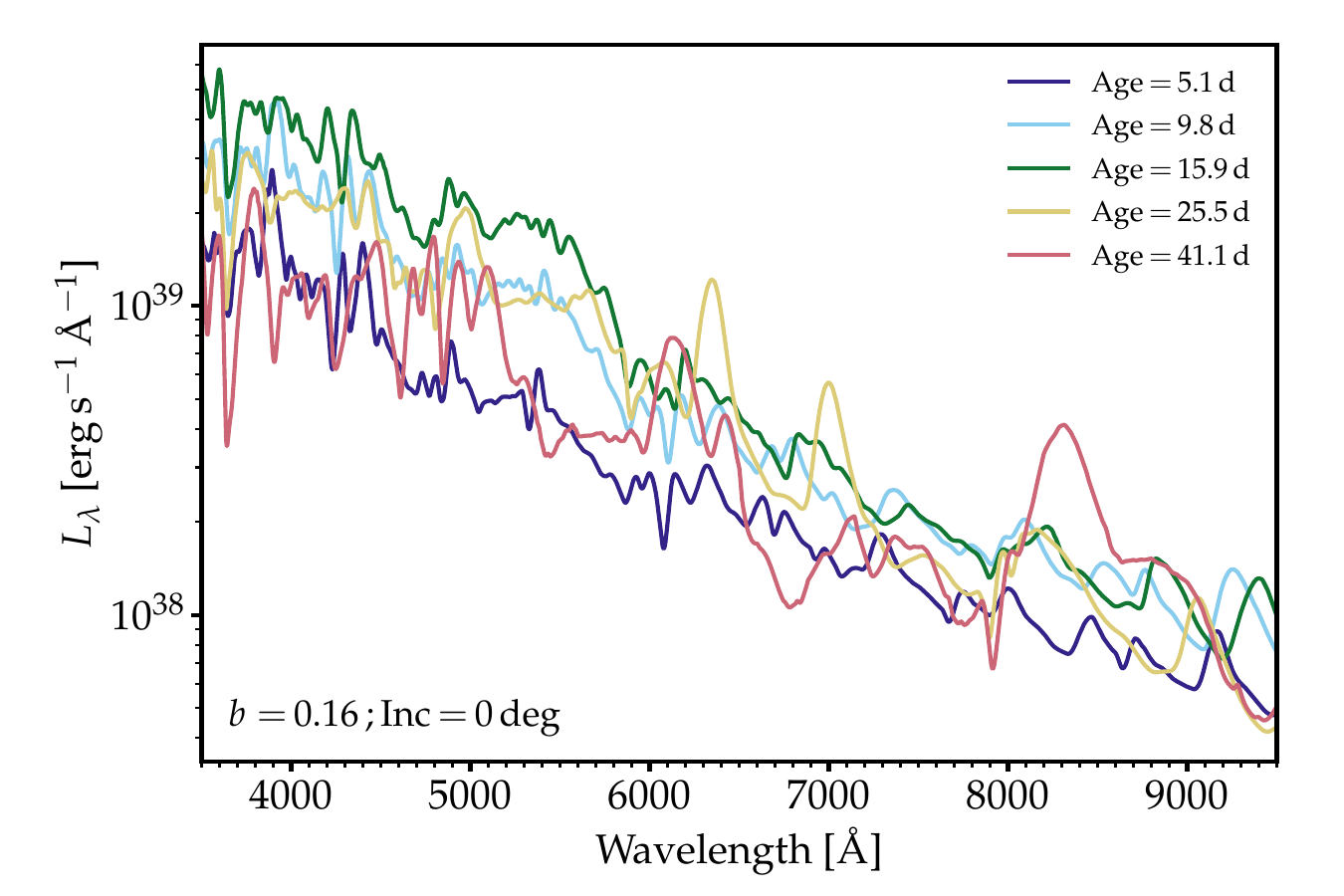}
    \includegraphics[width=0.49\textwidth]{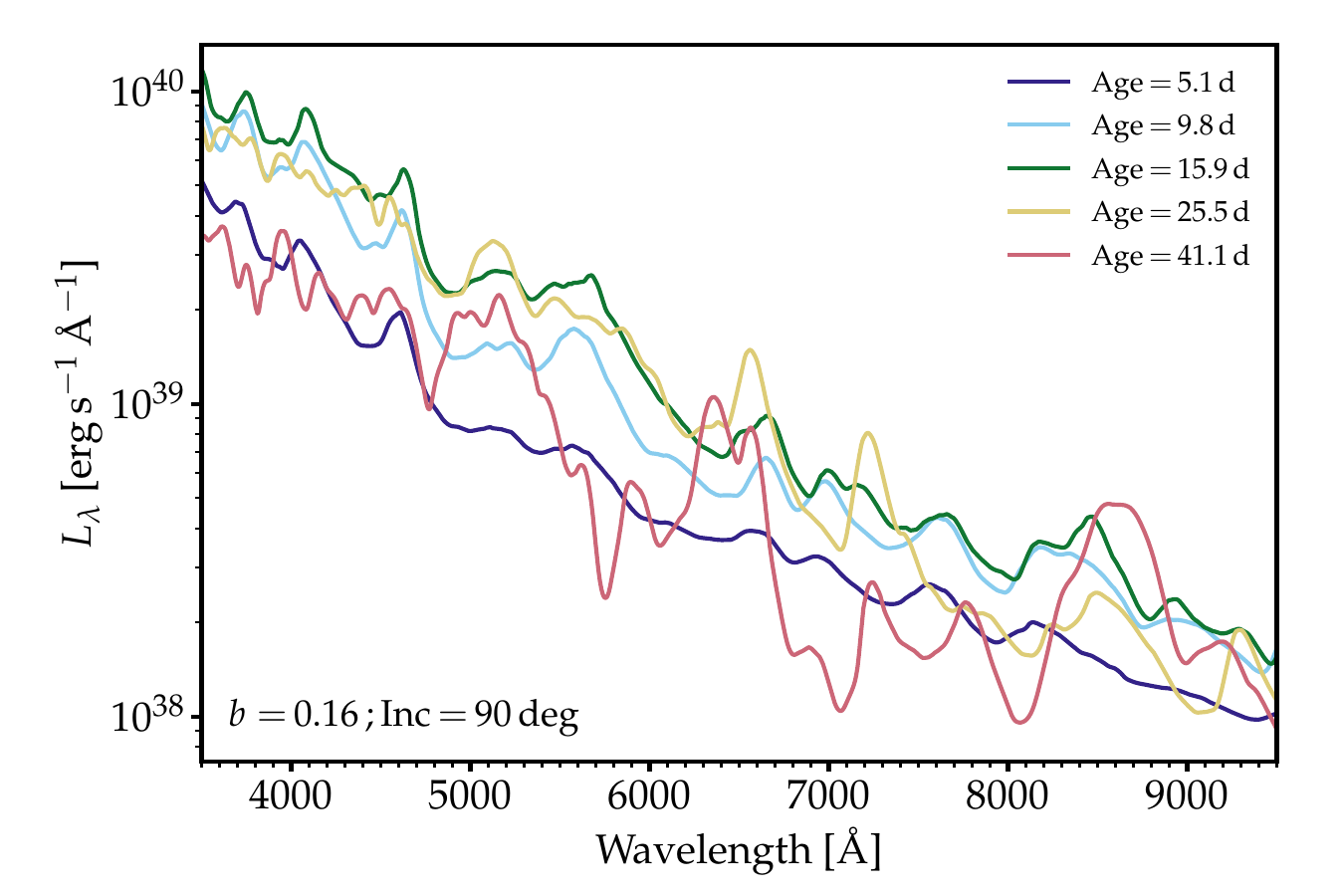}
    \includegraphics[width=0.49\textwidth]{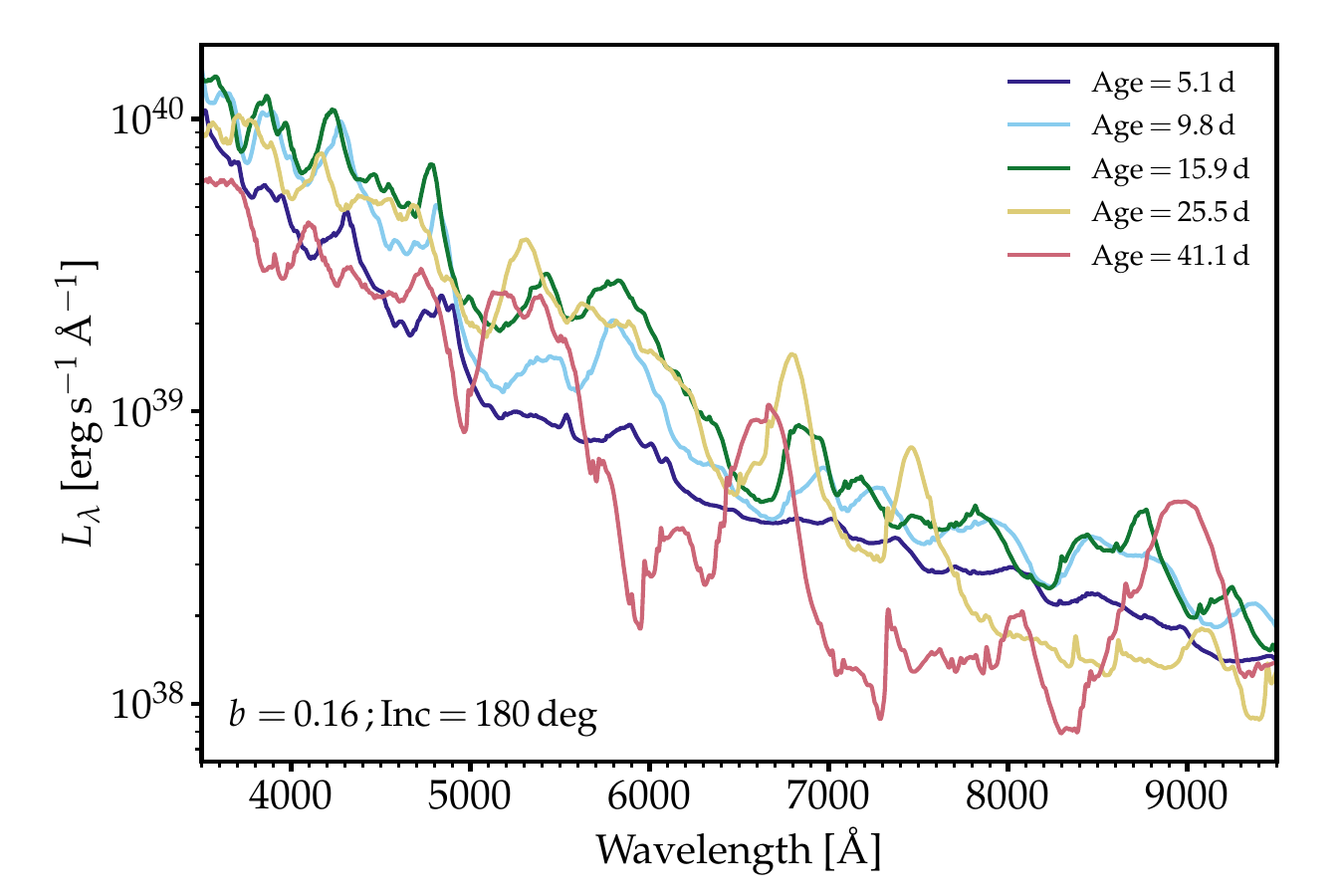}
    \caption{Spectral evolution in the optical for the 2D $b = 0.16$ model and for viewing angles of $0 \deg$ (top), $90 \deg$ (middle), and $180 \deg$ (bottom). The spectra change rapidly with wavelength at earlier times and lines become narrower with time. The spectral lines at $0 \deg$ and $180 \deg$ are blueshifted and redshifted, respectively (most evident at later times).}
    \label{fig:long_pol_b0p16_evol}
\end{figure}

\begin{figure}
    \includegraphics[width=0.49\textwidth]{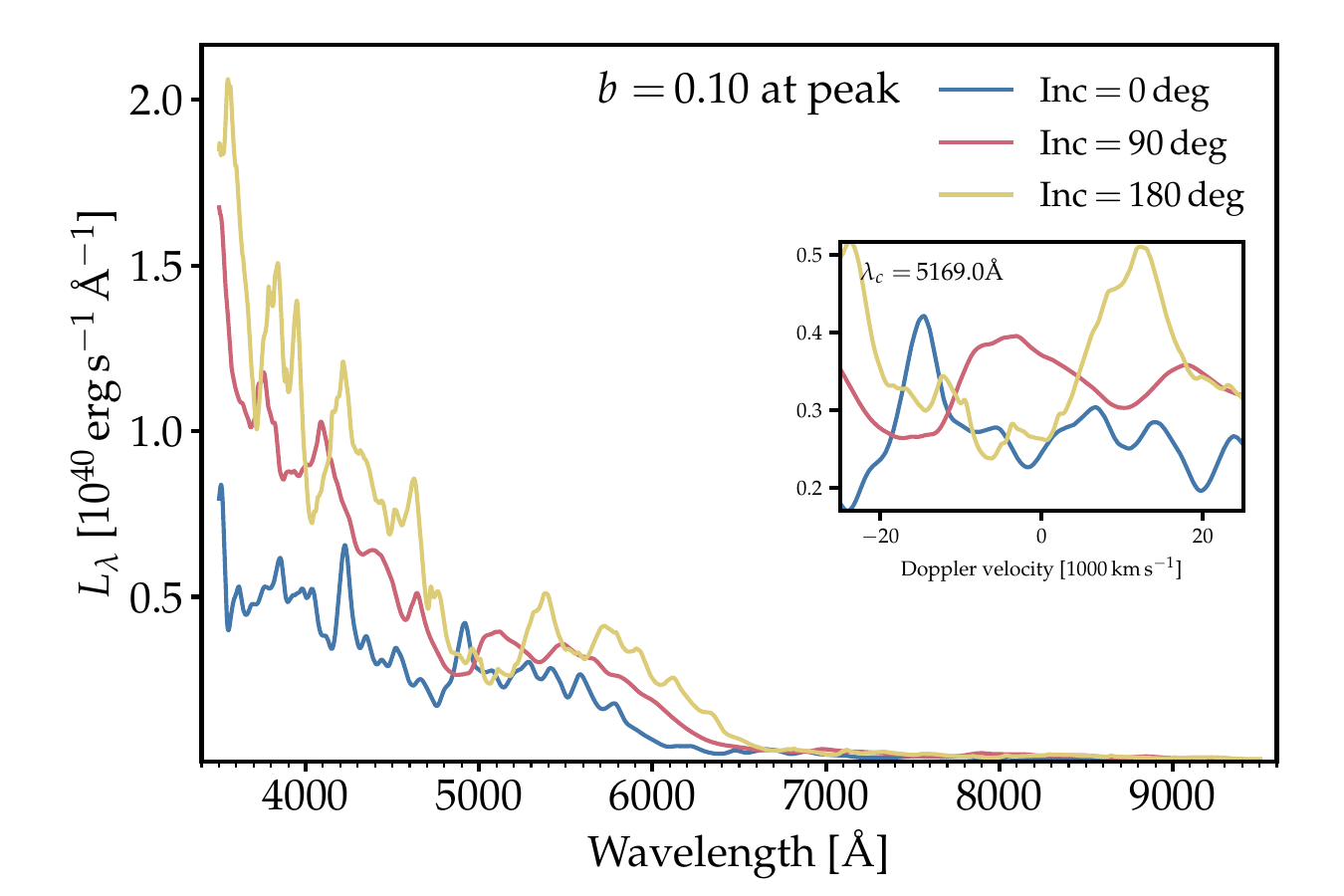}
    \includegraphics[width=0.49\textwidth]{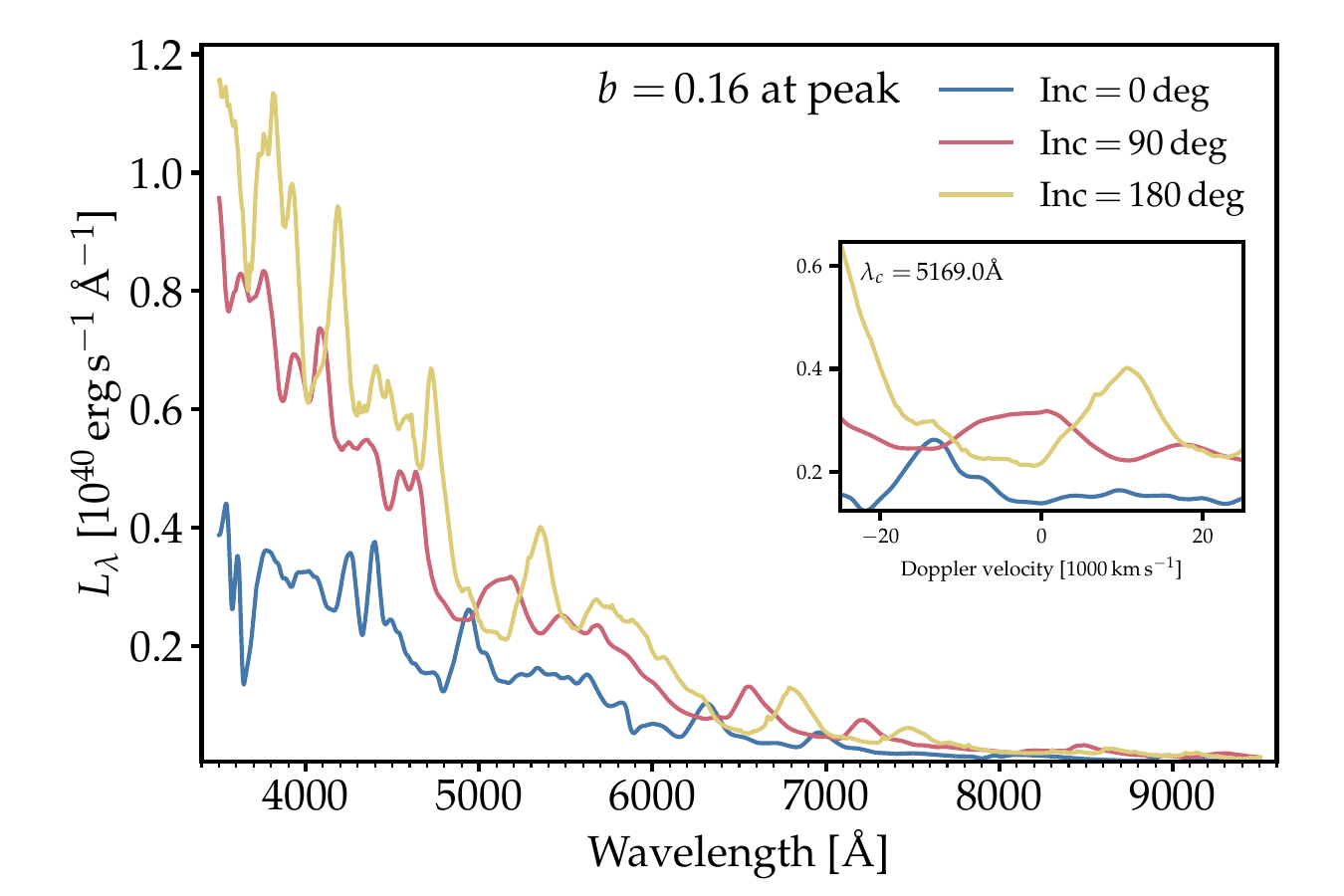}
    \includegraphics[width=0.49\textwidth]{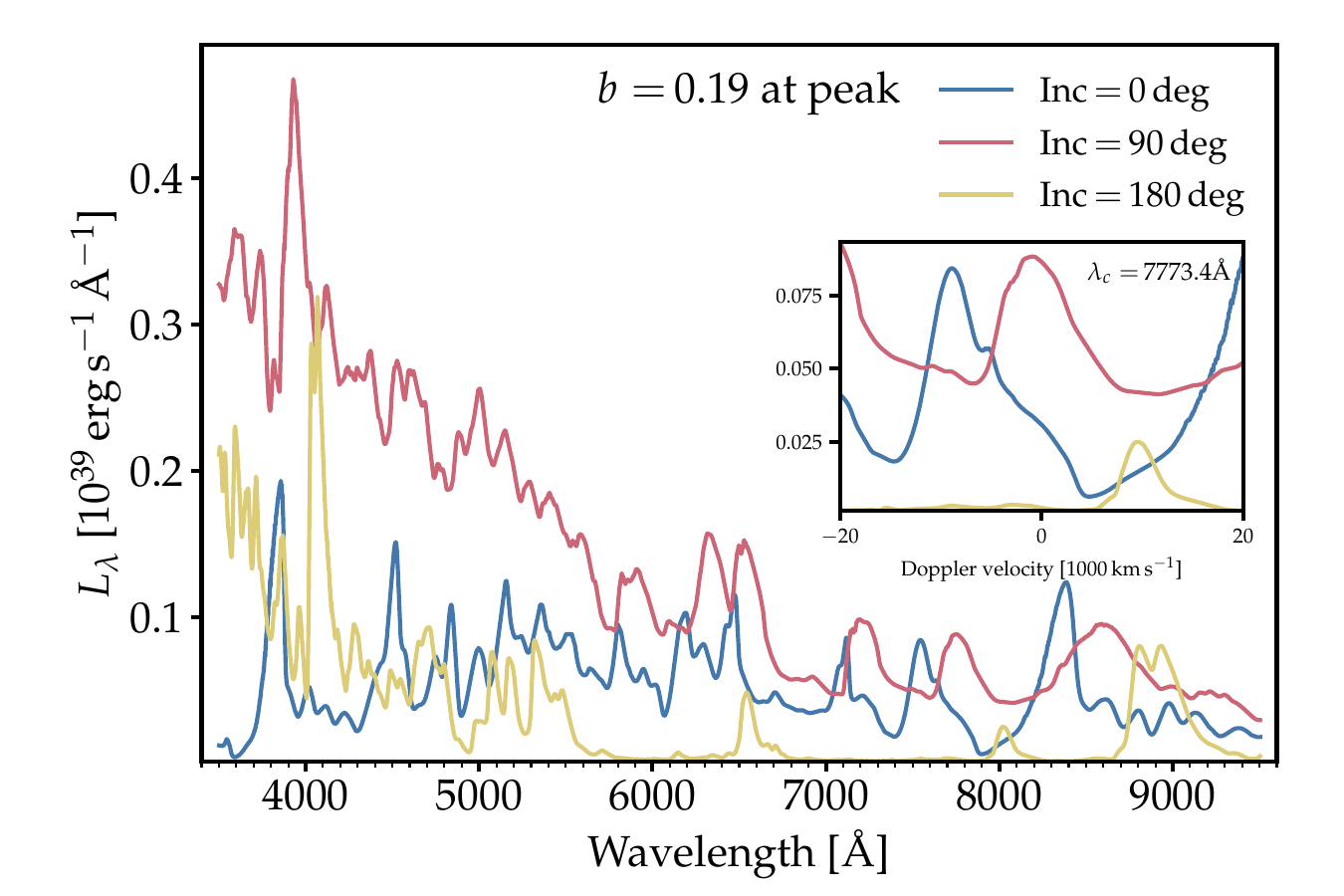}
    \caption{Optical spectra for models $b = 0.10$ (top), $b = 0.16$ (middle), and $b = 0.19$ (bottom) at the time of bolometric maximum (of the 1D model counterpart) for viewing angles of $0$, $90$, and $180 \deg$. The zoomed-in panels show wavelength shifts in the spectral lines that depend on viewing angle.}
    \label{fig:long_pol_max}
\end{figure}

\begin{figure}
    \includegraphics[width=0.49\textwidth]{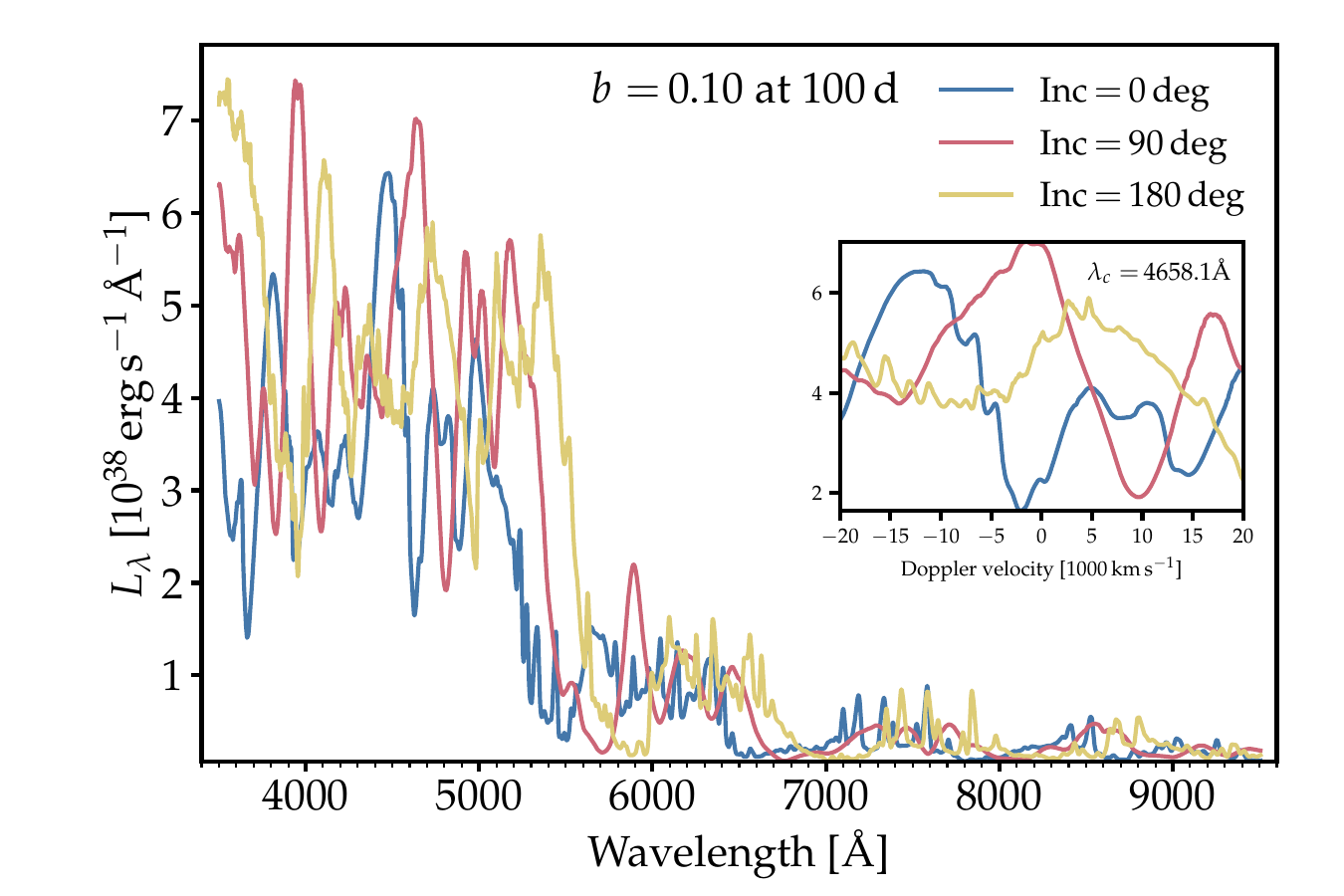}
    \includegraphics[width=0.49\textwidth]{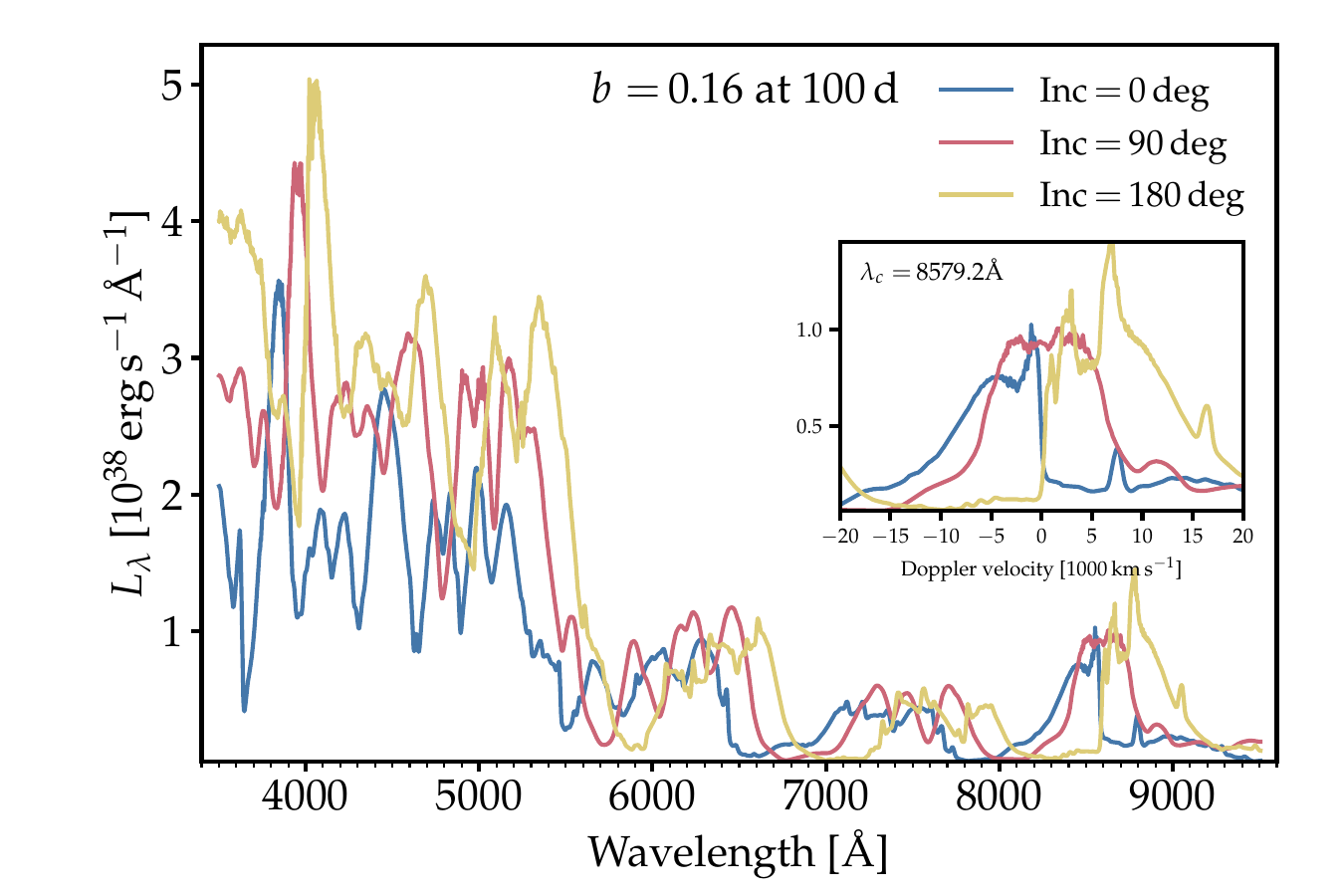}
    \includegraphics[width=0.49\textwidth]{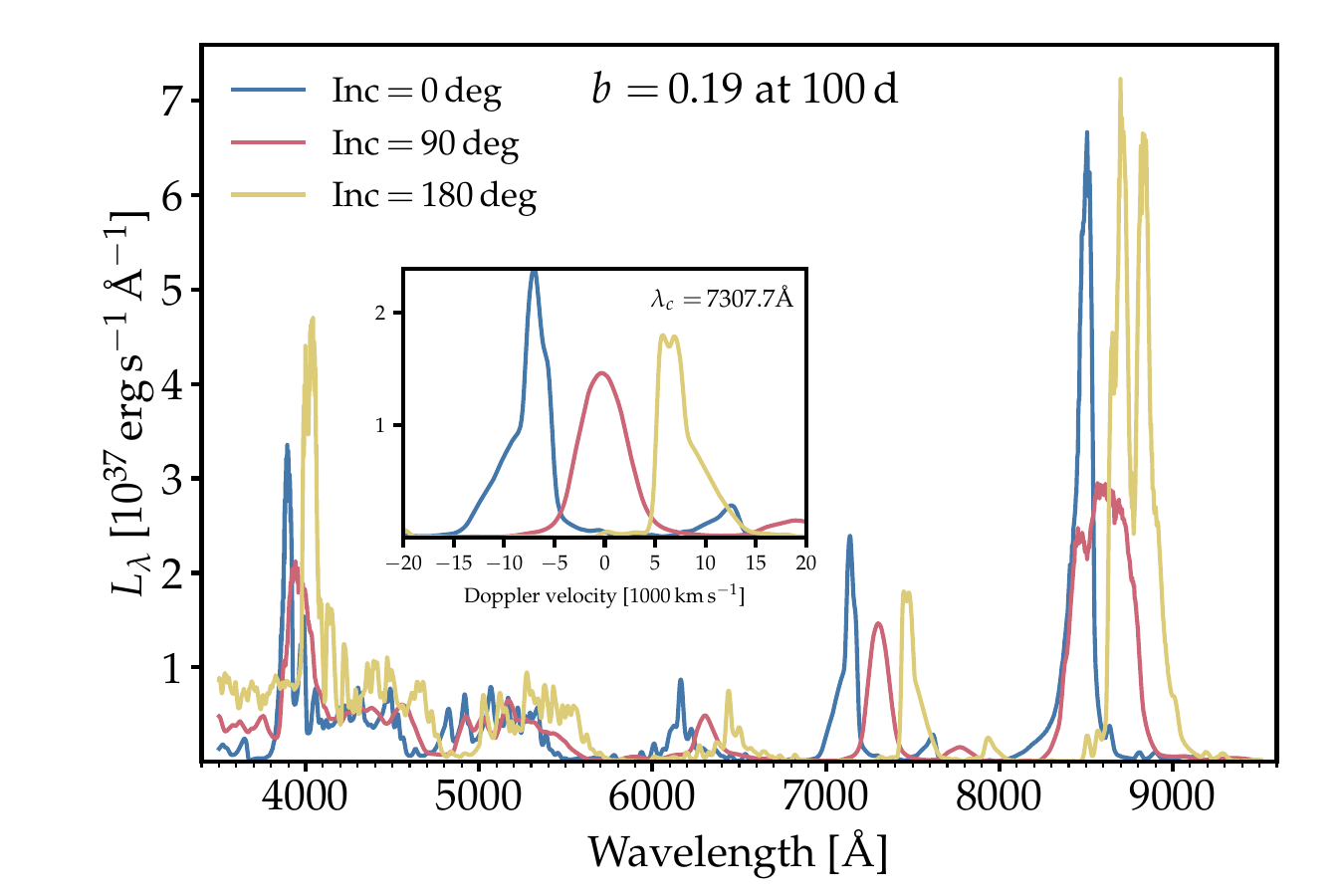}
    \caption{Optical spectra for models $b = 0.10$ (top), $b = 0.16$ (middle), and $b = 0.19$ (bottom) at $100 \dy$ and for viewing angles of $0$, $90$, and $180 \deg$. The zoomed-in panels show wavelength shifts in the spectral lines that depend on viewing angle.}
    \label{fig:long_pol_100d}
\end{figure}

Overall, and not surprisingly, these photometric properties are comparable to those of SNe Ia, be it observed ones or models thereof, that are characterized by the same range in \nifs\ mass. With the 2D simulations, only a limited spectral range was modeled, and thus no bolometric luminosity can be recovered from the \longpol\ calculations. We compared the $B$-band magnitudes instead. In the set of Chandrasekhar mass and sub-Chandrasekhar mass models of \citet[these simulations are a good set to compare to since they match for the most part the width-luminosity relation, as well as the photometric and spectral evolution, of SNe Ia]{2017MNRAS.470..157B}, the peak $B$-band magnitudes span the rough range from $-17$ to $-19$\,mag for \nifs\ masses in the range $\sim 0.1$ -- $0.5 \Msun$, with corresponding $B$-band rise times of $\sim 17 \dy$. Our set extends to much lower peak $B$-band magnitudes because it reaches down to very low \nifs\ masses (e.g., $0.007 \Msun$ with model $b = 0.19$). Our model rise times cluster around $20 \dy$, thus slightly larger but in rough agreement with theirs. A possible reason for this systematic shift may be the slower expansion of the ejecta that arises from the gravitational pull by the central BH, thus leading to a slightly larger ejecta diffusion time relative to standard SNe Ia. What is unique with SN-TDEs is the strong viewing angle dependence of these ejecta, leading to a variation in brightness for the same \nifs\ mass.

Figure \ref{fig:long_pol_b0p16_evol} shows the spectral evolution from about 5 to $40 \dy$ for the $b = 0.16$ model for a viewing angle of $0$, $90$, and $180 \deg$. This covers most of the photospheric phase and thus straddles the time of bolometric maximum. Hence, the spectra shift upwards until about $20 \dy$ before shifting downwards towards a lower power. For a viewing angle of $0 \deg$, one observes the debris along the direction of expansion. At the earliest time, the spectra show much structure and rapidly change with wavelength. This arises from the fact that we are essentially observing the troughs of what would in 1D appear as a fully developed P-Cygni profile with both absorption and a broad emission (see Figure \ref{fig:lbol_spec_at_max} or Figure \ref{fig:spec_all}) -- only the absorption part is essentially present for that viewing angle because the debris are confined within a narrow cone pointing towards the observer.  As time progresses, the photosphere recedes in the ejecta, lines become narrower but still exhibit a systematic blueshift. 

For a $90 \deg$ viewing angle, the ejecta expansion is perpendicular to the line of sight. At once, the observer captures emission from all ejecta depths but with a smaller range in projected velocity due to the finite angular extent of the cone of $\sim 30 \deg$ seen at a $90 \deg$ angle. 

For a $180 \deg$ viewing angle, the ejecta expansion is now away along the line of sight, and the emission arises preferentially from the inner, hot, and dense inner ejecta layers. All line emission and absorptions are consequently redshifted, although this is not easily noticed because of the myriad of contributing lines and their overlap. This Doppler shift is, however, evident when one compares one epoch for the three viewing angles. For example, a vertical line (as can be created by using the edge of the computer screen), clearly stands redward, along the center, or blueward of the Ca\two\ NIR emission for viewing angles of $0$, $90$, or $180 \deg$.

Figure \ref{fig:long_pol_max} illustrates the spectral properties of the $b = 0.10$, $0.16$, and $0.19$ models at bolometric maximum, but comparing the properties for the three viewing angles of $0$, $90$, and $180 \deg$ in each panel. The flux is usually greater for $90$ and $180 \deg$ angles since they reveal the inner, hotter ejecta layers, although this can be mitigated by the chemical stratification (e.g., if \nifs\ is present throughout the ejecta as in the $b = 0.10$ model or more confined to the inner ejecta as in the $b = 0.19$ model). Insets in the three panels of Figure \ref{fig:long_pol_max} show a zoom-in on the flux in Doppler velocity centered on the strong transitions of Fe\two\ at $5169 \ang$ or O\one\ at $7773.4 \ang$. There is evidently a shift from blue to red as the viewing angle increases, although even in this narrower spectral range, additional transitions contribute.

Figure \ref{fig:long_pol_100d} is a counterpart of Figure \ref{fig:long_pol_max} but now at $100 \dy$. At nebular conditions, the spectrum is dominated by a few strong emission lines, including gaps with no emission beyond $5500 \ang$, essentially associated with regions devoid of Fe\two\ lines. Here too, some insets are used to reveal the wavelength shifts in some strong emission lines, using as diagnostics the air wavelength of [Fe\three]\,$4658 \ang$, the $gf$-weighted\footnote{$gf$ refers to the oscillator strength of an atomic transition.} mean wavelength of the Ca\two\ NIR triplet, or that of \caiidoub.

The projected velocity of the emission is shifted by $5000$ -- $10000 \kmsinv$ as the viewing angle changes from $0$ to $180 \deg$. In addition, we clearly see the absence of emission from within $5000 \kmsinv$ for viewing angles of $0$ and $180 \deg$, which arises from the lack of material below about $5000 \kmsinv$ in all the SNe Ia TDEs simulated here. This is directly visible in the $b = 0.19$ model, but harder to identify in the $b = 0.16$ model in which we used the Ca\two\ NIR triplet (each component at $8498.0 \ang$, $8542.1 \ang$, and $8662.1 \ang$ is optically thick and affects the whole triplet out to many $1000 \kmsinv$) or the $b = 0.10$ model because \nifs\ is originally present throughout the ejecta (i.e., a large range of depths contribute line emission).

%%%%%%%%%%%%%%%%%%%%%%%%%%%%%%%%%%%%%%%%%%%%%%%%%%%%%%%%%%%%%%
\section{Discussion and conclusions} \label{sec:conclude}

We explored close TDEs of C/O WDs due to IMBHs using the 3D hydrodynamics code \arepo, including an on-the-fly NRN with 55-isotopes, and the 1D and 2D radiative transfer codes \cmfgen\ and \longpol, respectively. Setting the masses of the WD and the IMBH to $0.6 \Msun$ and $500 \Msun$, respectively, we studied the effect of the scaled encounter impact parameters $b = \rp/\rt$ (ranging from $0.10$ to $0.20$) on their outcomes. We ran the 3D hydrodynamics simulations until $\sim 500 \s$ following periapsis passage, after which we employed the 1D radiation hydrodynamics code \voned\ to run them further until homologous expansion of the ejecta was attained at $1 \dy$. These data were then input into the 1D and 2D radiation transfer codes to obtain their light curves and spectra in the photospheric ($\lesssim 40 \dy$) and nebular phases ($\sim 100 \dy$).

The most striking result of our WD TDE simulations is the strong dependence of the chemical composition and morphology of the ejecta on $b$. Stronger encounters (lower $b$) yield higher peak temperatures in the tidally compressed WDs, resulting in increased nuclear burning of \ctwelve\ and \osixteen\ into heavier isotopes. Consequently, while a wider encounter (e.g., $b = 0.20$) leads to a standard full TDE (the WD is completely destroyed) with negligible burning, a closer encounter (e.g., $b = 0.10$) results in an SN 1a TDE where a full TDE is supplemented by runaway nuclear burning. The fractions of \nifs\ produced in such SNe Ia TDEs range from $1\%$ when $b = 0.19$ to $82\%$ when $b = 0.10$. We note that the specific case of $b = 0.20$ in \citet{2009ApJ...695..404R}\footnote{Runs 8 and 9 of Table 1; same WD and IMBH masses} resulted in a SN explosion, while it is akin to a standard TDE with negligible nuclear burning in our simulations. This can be attributed to their relatively low resolution of $\sim 5\times10^5$ SPH particles constituting their WD, compared to our $6\times10^6$ cells. During our resolution tests, we observed that lower-resolution simulations had higher rates of nuclear fusion (see also Section \ref{sec:hydro}), likely due to the larger volumes of cells where ignition occurs.

Due to the energy injected into the ejecta from nuclear fusion reactions, the debris material spreads out more in SNe Ia TDEs compared to the thin tidal streams of standard TDEs. Thus, while $\sim 55$\% of WD material is unbound when $b = 0.20$ (similar to a standard TDE), $\sim 85\%$ of WD is unbound when $b = 0.10$ (significant nuclear burning). Moreover, due to the inherent asymmetry of the WD-IMBH encounter, most of the ejecta are present within a cone of $\sim 30\%$, resulting in very asymmetric SN explosions. We expect a similar dependence of nuclear burning rates on $b$ for different IMBH masses, although the exact rates and, consequently, massive isotope fractions would vary due to differences in tidal compression and effects of general relativity. Altering the WD masses (and hence their compositions), on the other hand, can lead to distinct outcomes because nuclear reactions in He WDs ($0.15 \Msun \lesssim \mwd \lesssim 0.45 \Msun$; e.g., \citealt{2013A&A...557A..19A}), O/Ne WDs ($1.05 \Msun \lesssim \mwd \lesssim 1.35 \Msun$; e.g., \citealt{2015MNRAS.446.2599D}) or even C/O WDs of different masses ($0.45 \Msun \lesssim \mwd \lesssim 1.05 \Msun$; e.g., \citealt{2010A&ARv..18..471A}) would play out quite differently. We leave these for future studies.

SNe Ia TDEs are most dramatically distinct from garden-variety SNe Ia in the extreme asymmetry of the debris and the presence of a central BH. Even if assumed dormant, the gravitational pull on the ejecta extends the dynamical phase (i.e., the debris moves ballistically only after a number of hours) and leads to a fallback that arves the inner ejecta below a few $1000 \kmsinv$. The debris asymmetry also leads to a dependence of brightness and colors with viewing angles, although this also depends on the \nifs\ mass and the chemical stratification. The most striking spectral signature of SNe Ia TDEs, best visible at nebular times, is the large velocity shifts of emission lines for viewing angles away from $90 \deg$, which arise from the strong ejecta asymmetry. This shift is aggravated by the cavity in the inner ejecta, resulting in a large displacement of the line emission (i.e., both a skewness and a shift of line emission). Interestingly, this has been observed in a fast-blue optical transient with systematically blueshifted and displaced H\one\ emission lines \citep{2024ApJ...977..162G} -- this event was interpreted as the TDE of an unevolved, low-mass star by an IMBH. 

In this work, we have held back from making a direct comparison of the observables from the radiative-transfer calculations and the current observed dataset of SNe Ia. One reason is that we are likely still awaiting a firm observation of an SN Ia TDE (see, however, \citealt{2025arXiv250925877L,2026MNRAS.546ag005E}). This work is also based on the TDE of a $0.6 \Msun$ C/O-WD, which is a very small mass relative to that inferred for an SN Ia (for example, such a light WD makes hardly any \nifs\ in the single- or double-degenerate scenario; see, for example, \citealt{2024A&A...683A..44M}). Another is the inconsistency in our approach for the 2D radiative transfer and the difficulty in evaluating the associated uncertainty in the predictions. While qualitatively sound, we expect systematic quantitative offsets at all photospheric epochs. At nebular times, our method and its predictions are more robust. Still, our enforcement of homogeneity within radial shells when computing both the 1D and 2D radiative transfer compromises the accuracy of the prediction. The reason is that such an enforced mixing alters the relative strength of coolants in the ejecta and thus the strength of spectral emission lines \citep{1989ApJ...343..323F, 2020A&A...642A..33D}. This should have little impact in models with a high \nifs\ mass fraction (e.g., models with $b \ge 0.15$), but it is responsible for the weakness of the \oidoub\ emission in the $b = 0.19$ model at $100 \dy$ (see bottom row of Figure \ref{fig:long_pol_100d}).

Future improvements will require using a better description of the spatial distribution of radioactive decay power within the debris, as well as the viewing angle dependence of the luminosity. Such multidimensional constraints could be computed using \sedona\ (\citealt{2006ApJ...651..366K}; as in \citealt{2016ApJ...819....3M}), feeding such constraints (e.g., the viewing-angle dependent energy deposition from radioactive decay or the viewing angle dependent bolometric luminosity) as inputs for the radiative transfer calculation with \cmfgen\ and \longpol, as done for the simulations of 2D double detonations in white dwarfs \citep{2025ApJ...987...54B}.

%%%%%%%%%%%%%%%%%%%%%%%%%%%%%%%%%%%%%%%%%%%%%%%%%%%%%%%%%%%%%%
\begin{acknowledgements}
PV thanks Bart Ripperda and Norman Murray for providing access to their computational time on Compute Canada’s \textit{Niagara} cluster. PV thanks CITA for access to the \textit{Sunnyvale} cluster. LD acknowledges support from the ESO Scientific Visitor Program for a visit to ESO-Garching during the summer of 2025. LD acknowledges access to the HPC resources of TGCC under the allocation 2024 -- A0170410554 on Irene-Rome made by GENCI, France. \\
PV and LD contributed equally to the paper, with them conducting, analyzing, and writing the hydrodynamics simulations and the radiative transfer calculations, respectively. All authors contributed to the project planning, analysis, and writing.
\end{acknowledgements}

%%%%%%%%%%%%%%%%%%%%%%%%%%%%%%%%%%%%%%%%%%%%%%%%%%%%%%%%%%%%%%
% WARNING
% Please note that we have included the references below in
% order to compile the document, but we ask you to:
%
% - use BibTeX with the regular commands:
  \bibliographystyle{aa} % style aa.bst
  \bibliography{TDE_SNe} % your references Yourfile.bib
% - join the .bib files when you upload your source files
%%%%%%%%%%%%%%%%%%%%%%%%%%%%%%%%%%%%%%%%%%%%%%%%%%%%%%%%%%%%%%

% %%%%%%%%%%%%%%%%%%%%%%%%%%%%%%%%%%%%%%%%%%%%%%%%%%%%%%%%%%%%%%%
% % Appendices must be placed after   \end{thebibliography}
% % They will be placed automatically on a new page.
% %%%%%%%%%%%%%%%%%%%%%%%%%%%%%%%%%%%%%%%%%%%%%%%%%%%%%%%%%%%%%%%
\begin{appendix}
% %%%%%%%%%%%%%%%%%%%%%%%%%%%%%%%%%%%%%%%%%%%%%%%%%%%%%%%%%%%%%%%

\onecolumn

\section{Density profile of the white dwarf}

\begin{figure*}[ht!]
    \centering
    \includegraphics[width=0.9\textwidth]{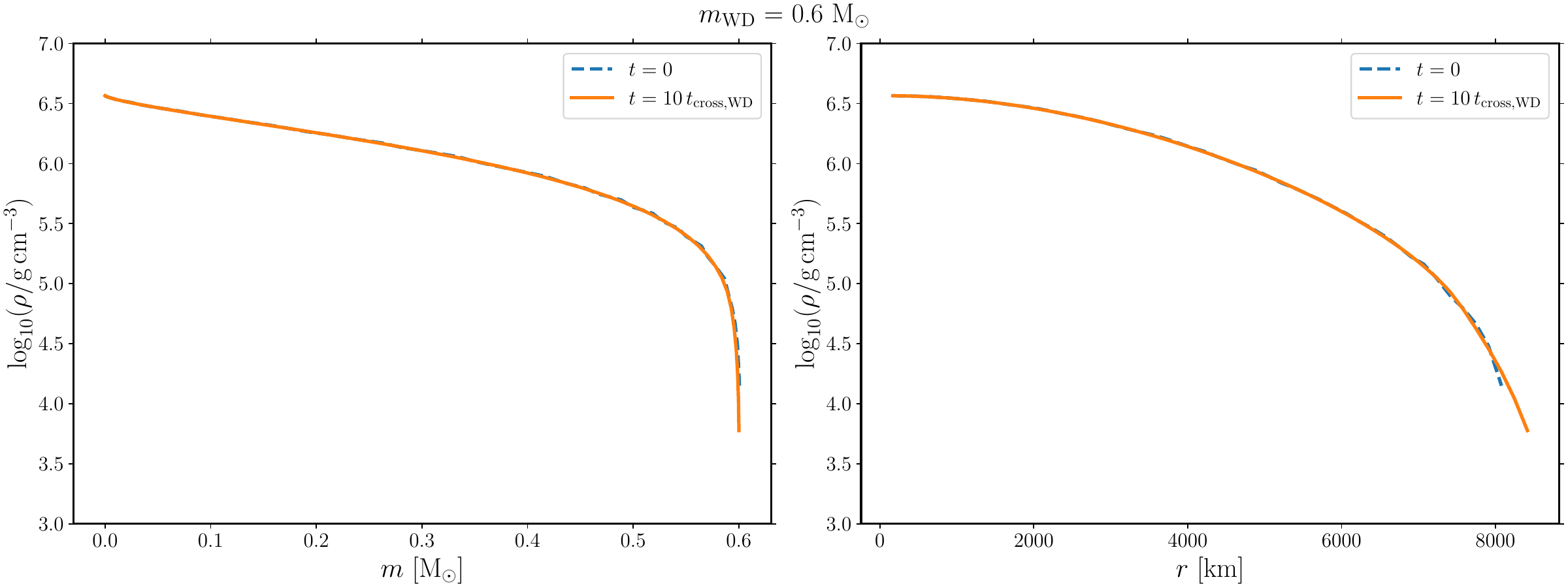}
    \caption{Density profiles of the \arepo-initialized $0.6 \Msun$ WD ($10^{-7}~\Msun/\mathrm{cell}$ resolution) as a function of mass (left) and radial (right) coordinates. The blue dashed and the orange solid lines denote the profiles initially (identical to 1D \mesa\ profile) and after 10 sound-crossing times ($\sim 127 \s$). The profiles overlap significantly.}
    \label{fig:density_wd}
\end{figure*}

\section{Kinetic energies of the unbound ejecta}

\begin{figure*}[ht!]
    \centering
    \includegraphics[width=0.24\textwidth]{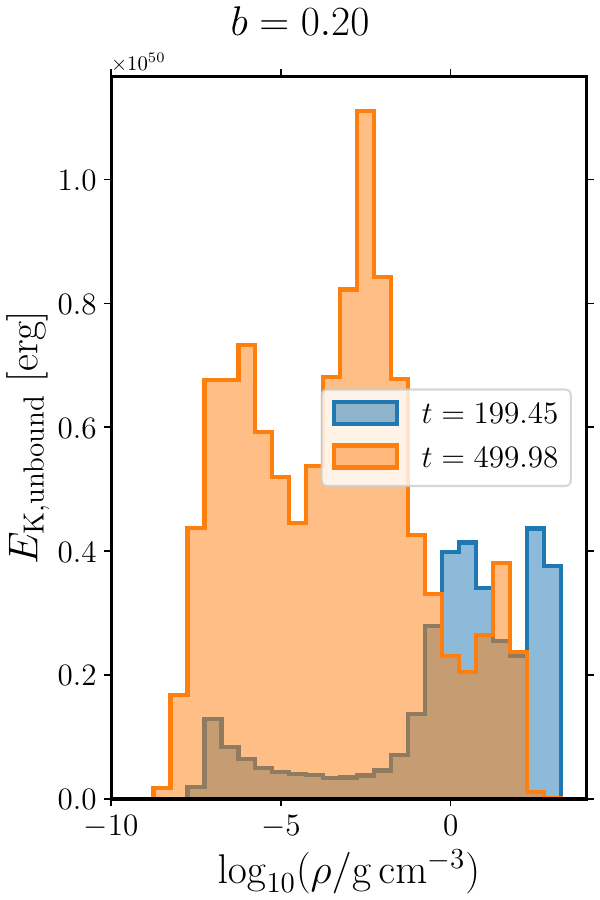}
    \includegraphics[width=0.24\textwidth]{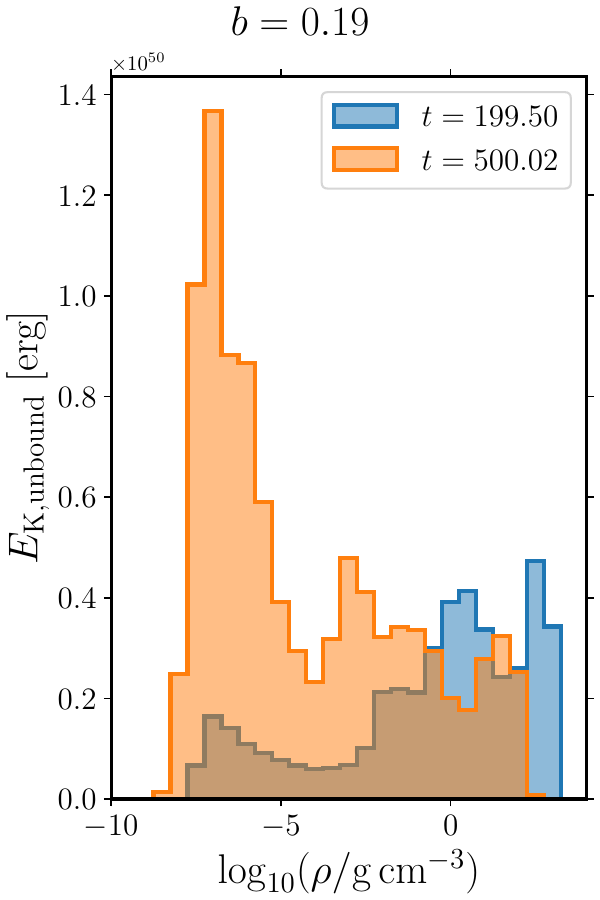}
    \includegraphics[width=0.24\textwidth]{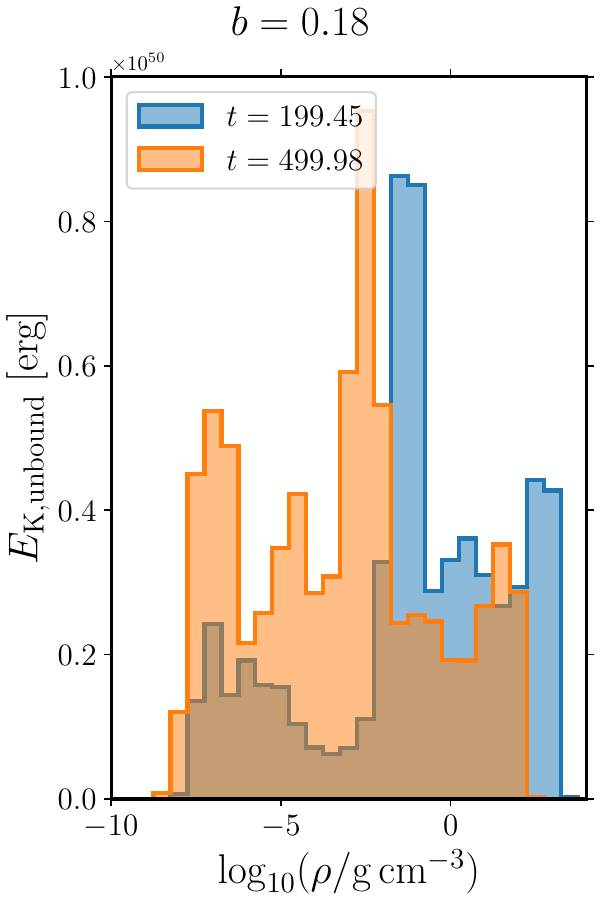}
    \includegraphics[width=0.24\textwidth]{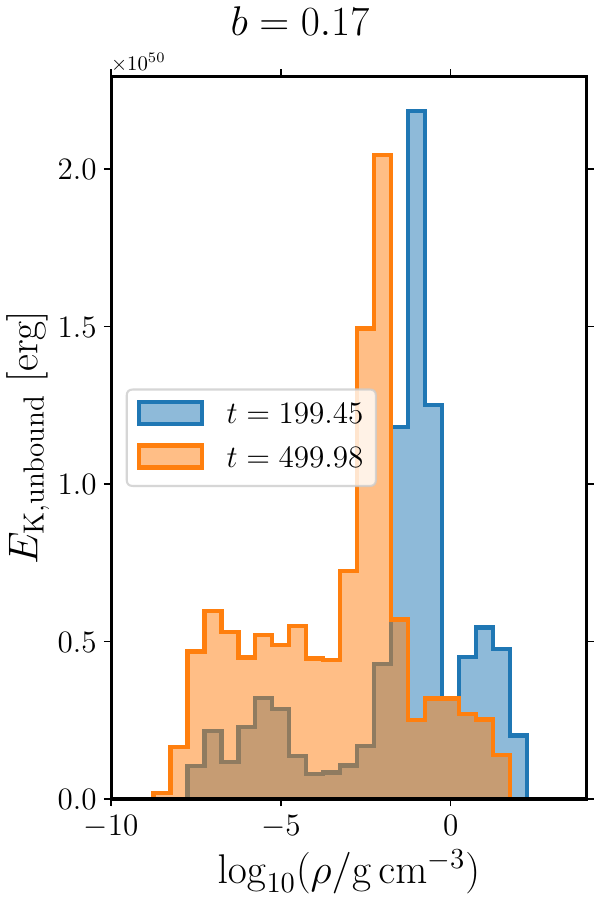}
    \caption{Kinetic energies $\EKunb$ of the unbound ejecta per 0.5 dex in density at $\sim 200 \s$ and $\sim 500 \s$, for $b$ ranging from $0.17$ to $0.20$. For $b = 0.20$ (and $b = 0.19$), $\EKunb$ at $\sim 500 \s$ is dominated by low-density, (very) high-velocity material, and is hence larger than at $\sim 200 \s$. However, as $b$ decreases (e.g., $b = 0.17$ and lower, not shown) and nuclear fusion becomes significant, there is only a slight change in $\EKunb$ with time owing to the decreased presence of such material. The peaks of the kinetic energies shift to lower densities with time due to the expansion of the ejecta.}
    \label{fig:kinE_unbound}
\end{figure*}

\twocolumn

\section{Latitudinal scaling in 2D radiative transfer}

\begin{figure}[ht!]
    \centering
    \includegraphics[width=0.5\textwidth]{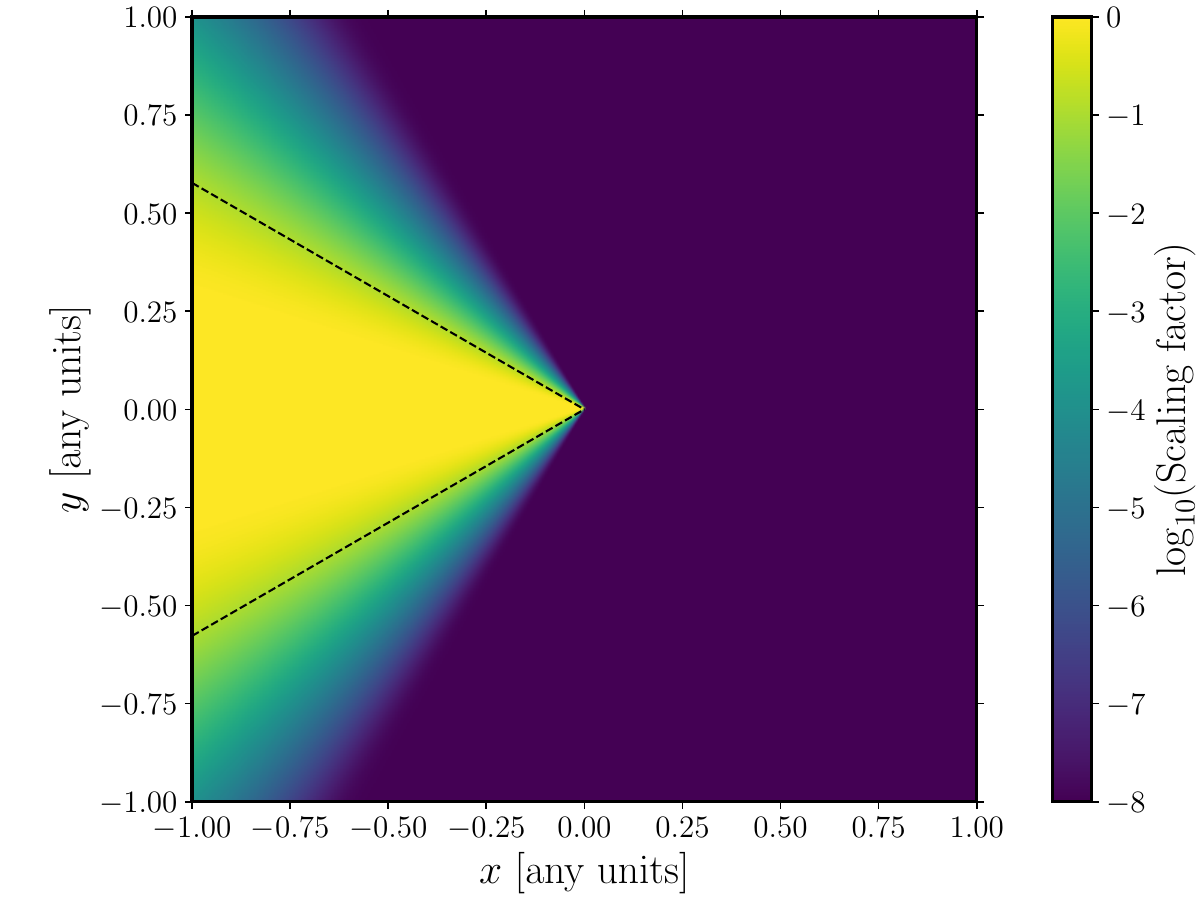}
    \caption{Latitudinal scaling used for the inputs (opacities $\chi$, emissivities $\eta$, and electron densities $N_e$) to the 2D radiative transfer code \longpol. The dashed lines represent the boundaries of a cone of $30^{\circ}$. The parameters $\alpha$ and $\beta$ set the steepness of scaling drop-off beyond the conical region and the gaussian scaling centered about the $-x$ direction, respectively. This scaling is performed to mimic the asymmetry of the WD TDE ejecta, shown in Figure \ref{fig:collage_quants}.}
    \label{fig:lat_scale}
\end{figure}

% \clearpage

\end{appendix}

\end{document}